\documentclass[reprint,pra,groupedaddress,longbibliography]{revtex4-1}

\usepackage{amsmath}
\usepackage{amssymb}
\usepackage{enumitem}
\usepackage{graphicx}
\usepackage{dcolumn}
\usepackage{multirow}
\usepackage{color}
\newcommand{\Hamiltonian}{\mathcal{H}}

\DeclareMathOperator\erf{erf}

\newcommand{\bra}[1]{\ensuremath{\langle#1|}}
\newcommand{\ket}[1]{\ensuremath{|{#1}\rangle}}

\newcommand{\red}[1]{\textcolor{black}{#1}}

\begin{document}

\title{Achieving High Fidelity Single Qubit Gates in a Strongly Driven Silicon Quantum Dot Hybrid Qubit}

\author{Yuan-Chi Yang}
\email[]{yang339@wisc.edu}
\affiliation{Department of Physics, University of Wisconsin-Madison, Madison, Wisconsin, 53706, USA}

\author{S. N. Coppersmith}
\email[]{snc@physics.wisc.edu}
\affiliation{Department of Physics, University of Wisconsin-Madison, Madison, Wisconsin, 53706, USA}

\author{Mark Friesen}
\email[]{friesen@physics.wisc.edu}
\affiliation{Department of Physics, University of Wisconsin-Madison, Madison, Wisconsin, 53706, USA}

\date{\today}

\begin{abstract}
Performing qubit gate operations as quickly as possible can be important to minimize the effects of decoherence. 
For resonant gates, this requires applying a strong ac drive.  However, strong driving can present control challenges by causing leakage to levels that lie outside the qubit subspace.  Strong driving can also present theoretical challenges because preferred tools such as the rotating wave approximation can break down, resulting in complex dynamics that are difficult to control.  
Here we analyze resonant $X$ rotations of a silicon quantum double dot hybrid qubit within a dressed-state formalism, obtaining results beyond the rotating wave approximation. 
We obtain analytic formulas for the optimum driving frequency and the Rabi frequency, which both are affected by strong driving. 
While the qubit states exhibit fast oscillations due to counter-rotating terms and leakage, we show that they can be suppressed to the point that gate fidelities above $99.99\%$ are possible, in the absence of decoherence. 
Hence decoherence mechanisms, rather than strong-driving effects, should represent the limiting factor for resonant-gate fidelities in quantum dot hybrid qubits.
\end{abstract}


\maketitle

\section{Introduction}
Microwaves have emerged as a key tool for manipulating quantum dot qubits via electron spin resonance~\cite{Koppens2006,Koppens2008,Veldhorst2014}, electric-dipole spin resonance~\cite{Nowack2007,PioroLadriere2008,NadjPerge2010,Kawakami2014}, or resonantly driven gates in charge qubits~\cite{Kim2015}, singlet-triplet qubits~\cite{Shulman2014}, exchange-only qubits~\cite{Medford2013}, and quantum dot hybrid qubits~\cite{KimWardSimmonsEtAl2015,2016arXiv161104945T}.
By adapting techniques used on atomic and superconducting qubits~\cite{Nakamura2001}, such gates provide flexibility, for example, via phase control of the rotation axes.
Microwave driving can also protect against low-frequency charge noise~\cite{Yan2013,Jing2014}, which is a dominant source of dephasing in quantum dot qubits~\cite{Dial2013}.
Moreover, ac driving allows us to continually center the tuning at an optimal working point or sweet spot~\cite{Vion2002}, which provides additional noise protection~\cite{Medford2013,Kim2015,KimWardSimmonsEtAl2015,Wong2016}.

To improve gate fidelities further, we must perform the gates quickly and accurately, suggesting that we employ large ac driving amplitudes.
Since quantum dots are highly tunable, it is typically easy to enter the strong driving regime, where the Rabi frequency approaches the resonant frequency.
Under such conditions, it has long been known that complicated dynamics can occur (e.g., fast beating), and that the resonant frequency can become a function of the driving amplitude (the Bloch-Siegert shift~\cite{Bloch1940,Shirley1965}).
Both effects arise from the counter-rotating term that is present in a harmonic, oscillatory driving field, but which is ignored in the rotating wave approximation (RWA)~\cite{DressedAtom}.
Additionally, if nonqubit leakage states are present, they may be excited through strong driving~\cite{Motzoi2009}.
While such fast beating and leakage effects are coherent, they can present challenges for qubit control and ultimately reduce the quantum gate fidelity.

Here, we theoretically study strong driving effects in quantum dot hybrid qubits~\cite{Shi2012,Koh2012,KimShiSimmonsEtAl2014,Ferraro2014,Cao2016}, and we propose methods for improving the gate fidelity while maintaining high gate speeds.
In this system, there is a nearby leakage state, and we show that both fast oscillations and leakage can be suppressed by employing simple, shaped microwave pulse envelopes~\cite{Motzoi2009,Gambetta2011,Motzoi2013}.
Moreover, we show that the requirements for pulse shaping are not strict, and that the resulting gate fidelities can be greater than 99.99\% for gate times shorter than 1~ns.
Although we do not consider decoherence effects in this work, our results suggest that decoherence, not control errors, should remain the leading challenge for gate fidelities in the foreseeable future.

Several methods have been developed for moving beyond the RWA and characterizing the Bloch-Siegert shift.
For a system like a double dot coherently coupled to a microwave resonator~\cite{Frey2012,Petersson2012,Stockklauser2015,Mi2017}, it is essential to use a dressed-state model~\cite{CohenTannoudji1973} in which the qubit electron(s) and the resonator photon(s) are both treated quantum mechanically~\cite{Childress2004,Burkard2006,Trif2008,Hu2012,TaylorUnp}.
If, instead, the quantum dot is driven via a classical field, one may employ either a semiclassical~[\onlinecite{Song2016},\red{\onlinecite{PhysRevA.86.023831},\onlinecite{Lu2016}}] or a fully quantum model.
Both approaches are capable of describing corrections to the RWA, and have been used to describe strong driving in superconducting qubit systems~\cite{Oliver2005,Wilson2007,FornDiaz2010,Tuorila2010}.
When the qubit is driven through an energy level anticrossing, it is common to analyze the dynamics~\cite{Shevchenko2010} using Landau-Zener-St\"{u}cklberg (LZS) theory~\cite{Landau1932,Zener1932,Stuckelberg1932}; this is also a preferred method for investigating other strong-driving effects such as multiphoton resonances~\cite{Laird2009,NadjPerge2012,Stehlik2014,Scarlino2015}.
For the quantum dot hybrid qubit, it is common to perform gate operations away from the level anticrossing~\cite{KimWardSimmonsEtAl2015}, suggesting that LZS theory may not be optimal for describing these dynamics.
Moreover, LZS theory does not typically incorporate leakage states.
Here, we therefore develop a dressed-state model of a strongly driven quantum dot hybrid qubit.
Our approach allows us to treat strong driving effects perturbatively, up to arbitrary order in the driving strength, and it allows us to derive simple analytical formulas for the qubit dynamics.
The formalism naturally describes the oscillations caused by counter-rotating terms as well as leakage. 
Our analytical results indicate that the amplitude of the fast oscillations in the qubit dynamics is proportional to the driving amplitude, to leading order, suggesting a direct trade-off between fast operations and gate errors.  
We go on to show that pulse shaping greatly ameliorates these trade-offs. 
Our numerical simulations of qubit dynamics allow us to validate our analytical formulas, and enable us to compare the fidelities of several different gating methods, including pulse shaping.

The paper is organized as follows.
We first introduce our system, the microwave-driven silicon double quantum dot hybrid qubit, in Sec.~\ref{Sec:QuantumDotHybridQubit}. 
We also outline our calculation method based on a dressed-state formalism.
In Sec.~\ref{sec:results}, we apply this method to the quantum dot hybrid qubit and describe the results. 
In Sec.~\ref{Sec:PulseShape}, we consider several pulse-shaping protocols, and describe their effect on the gate fidelity.  
Our main conclusions are presented in Sec.~\ref{Sec:conclusion}.
Several appendices contain the technical details of our analysis.
Appendix~\ref{Sec:EffectiveTwoLevel} provides details about the transformation from a three-state model of the quantum dot hybrid qubit to an effective two-state model.
Appendix~\ref{Sec:DressedStateDetail} provides details of our dressed-state formalism.
As an example, we apply the method to a two-level system, obtaining simple analytical corrections going beyond the RWA.
Appendix~\ref{Sec:HybridQubitDetail} provides details of the dressed-state theory of the three-level quantum dot hybrid qubit. 
In Appendix~\ref{Sec:PulseShapeDetuning} we describe our results when the drive is applied to the detuning parameter.
(In the main text, we mainly consider the case of tunnel coupling driving.)
Finally, Appendix~\ref{Sec:XPI2} describes our results for ac-driven $X_{\pi /2}$ gates, in contrast to the $X_\pi$ gates considered in the main text.  

\section{Quantum Dot Hybrid Qubit \label{Sec:QuantumDotHybridQubit}}
Here, we review our theoretical model for the quantum dot hybrid qubit in the absence of decoherence from the environment.  
The qubit is comprised of three electrons in a double quantum dot with total spin quantum numbers $S=1/2$ and $S_z=-1/2$.
For the operating regime of interest, we consider the three-dimensional (3D) basis $\ket{\cdot S}\equiv \ket{\downarrow\! S}$, 
$\ket{\cdot T}  \equiv \sqrt{1/3} \ket{\downarrow\! T_0} - \sqrt{2/3} \ket{\uparrow\! T_-}$, 
and $\ket{S\cdot} \equiv \ket{S\! \downarrow}$, 
where $\ket{\cdot}$ denotes a dot with one electron, and $|S\rangle=(\ket{\uparrow\downarrow}-\ket{\downarrow\uparrow})/\sqrt{2}$, $|T_0\rangle=(\ket{\uparrow\downarrow}+\ket{\downarrow\uparrow})/\sqrt{2}$, and $|T_-\rangle=\ket{\downarrow\downarrow}$ denote the spin states of dots with two electrons
\cite{Shi2012,Koh2012}.
In this basis, the Hamiltonian is given by
\begin{equation}
\Hamiltonian = 
\begin{pmatrix}
-\frac{\varepsilon}{2} & 0 & \Delta_1 \\
0 & -\frac{\varepsilon}{2} + E_\text{ST} & -\Delta_2 \\
\Delta_1 & -\Delta_2 & \frac{\varepsilon}{2}
\end{pmatrix},
\label{Eq:Hamiltonian}
\end{equation}
where $\varepsilon$ is the detuning between the dots, 
$E_\text{ST}$ is the singlet-triplet energy splitting of the doubly occupied dot, 
and $\Delta_1$ ($\Delta_2$) are the tunnel couplings between the states $\ket{S\cdot}$ and $\ket{\cdot S}$ ($\ket{\cdot T}$)~\cite{Koh03122013}. 
The two lowest energy states $\ket{0}$ and $\ket{1}$ comprise the qubit, while the high-energy state $\ket{L}$ is a leakage state.. 

\begin{figure*}[t]
\includegraphics[width=6in]{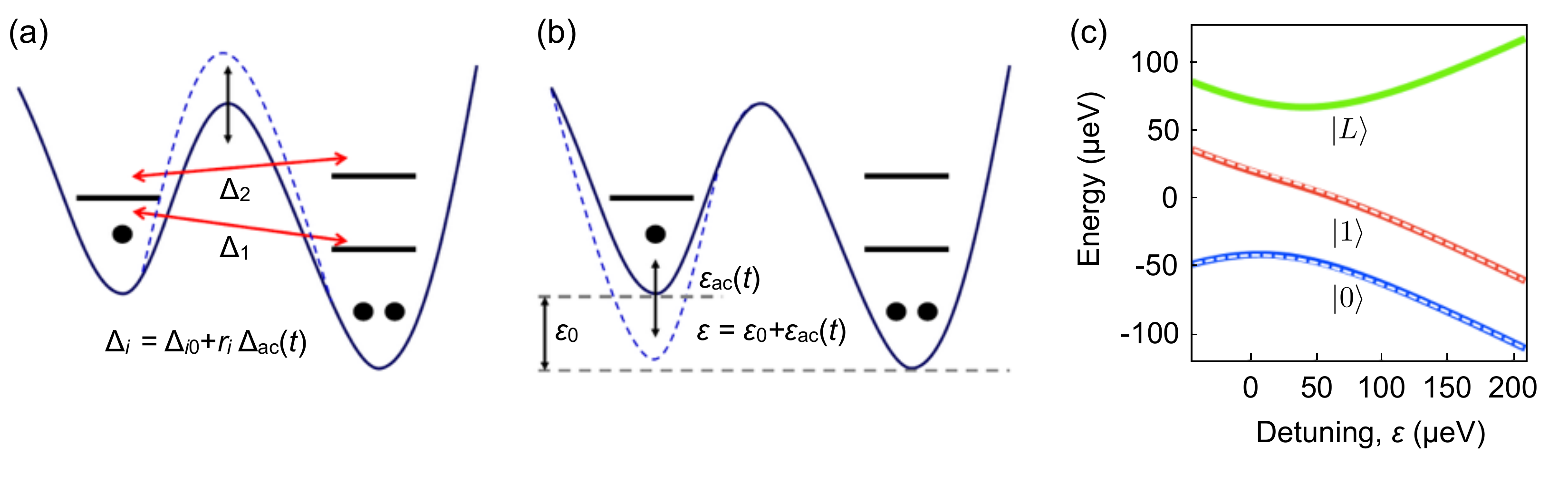}
\caption{
\label{Fig:HybridQubit}
Control of a quantum dot hybrid qubit.
(a),(b) Double quantum dots with three electrons, depicted here in their $(1,2)$ charge configurations.
In this arrangement, the low-lying energy levels, depicted in the right-hand dots, correspond to singlet-like ($\ket{\cdot S}$) and triplet-like ($\ket{\cdot T}$) spin states, where $S$ and $T$ refer to the two-electron dots~\cite{Shi2012,Koh2012}.
The $(2,1)$ charge configuration has only one low-lying energy level ($\ket{S\cdot}$), as depicted in the left-hand dots.
(a) illustrates control of the tunnel couplings $\Delta_1$ and $\Delta_2$, between $\ket{S\cdot}$, and $\ket{\cdot S}$ or $\ket{\cdot T}$, respectively.
(b) illustrates control of the detuning parameter $\varepsilon$, corresponding to the energy difference between the left and right quantum dots.
The parameters $\Delta_1$, $\Delta_2$, and $\varepsilon$ are all controlled by voltages applied to the double-dot top gates~\cite{KimWardSimmonsEtAl2015}.
Resonant gates are implemented by adding an ac drive to either the tunnel couplings or the detuning.
(c) A typical energy level diagram for the quantum dot hybrid qubit, as a function of $\varepsilon$, obtained by diagonalizing Eq.~(\ref{Eq:Hamiltonian}), using parameter values $E_\text{ST}/h = 12$~GHz and $\Delta_1=\Delta_2 = 0.7\,E_\text{ST}$~\cite{KimShiSimmonsEtAl2014}.
Here, the lowest two levels (red and blue) correspond to the qubit subspace, while the highest level (green) corresponds to a leakage state.
The effective two-level system of the qubit, derived in Eq.~(\ref{Eq:Heff}), is indicated by white dashed lines that overlay the qubit levels.}
\end{figure*}

In the absence of driving, a Schrieffer-Wolf transformation can be used to write an effective two-state Hamiltonian for the system, as described in Appendix~A.
Some typical energy level diagrams obtained using this method are plotted with white dashed lines in Fig.~\ref{Fig:HybridQubit}(c).
For experimental applications, we are often interested in the ``far-detuned" regime, $\varepsilon\gg E_\text{ST},\Delta_1,\Delta_2$, where the quantum dot hybrid qubit has a spin-like character.
In that case, we make use of the small parameters $E_\text{ST}/\varepsilon$, $\Delta_1/\varepsilon$, and $\Delta_2/\varepsilon$ to obtain simpler expressions for the transformation that diagonalizes $\Hamiltonian$, as described in the Appendix~\ref{Sec:HybridQubitDetail}.

For large detunings, the charge configuration of the double dot is approximately given by (1,2), which refers to the occupations of the left and right dots respectively, and the energy splitting between $|0\rangle$ and $|1\rangle$ is nearly independent of the detuning over an extended detuning range.
Such ``sweet spots" are protected from energy fluctuations and dephasing caused by charge noise, thus enabling long coherence times~\cite{KimShiSimmonsEtAl2014,KimWardSimmonsEtAl2015}.
The width of the sweet spot is maximized when $\Delta_1 \simeq \Delta_2\simeq 0.7E_\text{ST}$, since then the energy level anticrossings are closely spaced and the level repulsions induced by the tunnel couplings are nearly equal \cite{Wong2016}.  
Here, we focus on this optimal regime for the control parameters.

\subsection{Time-Dependent Hamiltonian:  The Semiclassical Approach}\label{sec:semi}
We consider two different schemes for ac driving \cite{Koh03122013}, as indicated in Figs.~\ref{Fig:HybridQubit}(a) and (b).
In the first scheme, we modulate the tunnel couplings, $\Delta_i = \Delta_{i0} + r_i \Delta_\text{ac}(t)$, where $i=1,2$.
The ac drive, $\Delta_\text{ac}$, is achieved by applying a microwave voltage signal to one of the top gates~\cite{KimWardSimmonsEtAl2015}.  
It is reasonable to assume that the ac signal drives both $\Delta_1$ and $\Delta_2$, although they may be affected differently, which we take into account through the variable $r_i$.
In the second scheme, we modulate the detuning, $\varepsilon = \varepsilon_{0} + \varepsilon_\text{ac}(t)$.
To simplify the discussion later, it is convenient to express all the driving functions as $u(t) = A \cos (2 \pi f\!t)$ where $u = \Delta_\text{ac}$ or $\varepsilon_\text{ac}$, $A$ is the respective driving amplitude, and $f$ is the driving frequency. 

Up to this point, our qubit Hamiltonian can be viewed as semiclassical, since the driving occurs through a time-dependent control parameter.
The resulting Hamiltonian is given by
\begin{equation}
\Hamiltonian_\text{semi} = \Hamiltonian_0 + V \cos (2 \pi f\!t) ,
\label{Eq:SemiHamiltonian}
\end{equation}
where $\Hamiltonian_0$ is obtained from Eq.~(\ref{Eq:Hamiltonian}) when $A=0$, and $V$ represents a driving term, which is proportional to $A$.
Note here that $\Hamiltonian_0$ and $V$ are both $3 \times 3$ matrices, $\Hamiltonian_0$ is constant, and the form of $V$ depends on which control parameter is driven. 

At this point, it is tempting to apply the analytical block-diagonalization procedure to $\Hamiltonian_0$ described in Appendix~A to construct an effective, time-dependent 2D Hamiltonian for the qubit subspace, and then study the dynamical evolution within this subspace.
This procedure is useful in the limit of small driving (for example, Fig.~\ref{Fig:HybridQubit}(c) shows the energy levels of the undriven system are well-described by the result of this analysis), but in the presence of strong driving it is flawed, because the unitary transformation that block-diagonalizes $\Hamiltonian_0$ does not block-diagonalize $V$.
Indeed, the leakage state, which lies outside the qubit subspace, plays an essential role in the dynamical evolution which cannot be modeled as a simple effective exchange interaction between the qubit levels.
Such leakage dynamics can be ignored in the weak-driving regime, but not the strong-driving regime.

\subsection{Dressed-State Hamiltonian: The Quantum Approach}\label{Sec:DressedStateFormalism}
While it is possible to analyze the semiclassical Hamiltoniann~(\ref{Eq:SemiHamiltonian}) with time-periodic driving using Floquet theory~\cite{Grifoni1998,doi:10.1063/1.445483,0022-3700-17-10-015,doi:10.1021/j150665a002,Shirley1965}, here we take the alternate approach of describing the electromagnetic field quantum mechanically.
Such methods were originally developed to describe the resonant interactions between atoms and photons, for example, in the form of laser or microwave driving fields~\cite{CohenTannoudji1973,Tannoudji1977,Tannoudji1985,DressedAtom}.   

We now develop a dressed-state formalism to describe the microwave driving of the quantum dot hybrid qubit. 
The first step is to note that our semiclassical expression for the Hamiltonian in Eq.~(\ref{Eq:SemiHamiltonian}) does not explicitly include a photonic driving field.
We now introduce such a photon field in its second quantized form:  $\Hamiltonian_\text{ph}= I_\text{dot} \otimes (h\!f a^{\dagger} a)$.
Here, $I_\text{dot}$ is the identity matrix acting on the double dot, $f$ is the microwave driving frequency, and $a^\dagger$ and $a$ are photon creation and annihilation operators.
Note that $I_\text{dot}$ has the same dimensionality as the dot Hamiltonian $\Hamiltonian_0$, and is 3D for the quantum dot hybrid qubit.
The dot Hamiltonian can similarly be written as $\Hamiltonian_\text{dot}=\Hamiltonian_0\otimes I_\text{ph}$, where we express $\Hamiltonian_0=\text{diag}[E_0,E_1,E_L]$ in the quantum dot hybrid qubit eigenbasis $\{\ket{i}\}$ $(i=0,1,L)$.
It is convenient to expand the uncoupled Hamiltonians in terms of the ``bare-state" basis $\{\ket{i,n}=\ket{i}\otimes\ket{n}\}$, where $\ket{n}$ represents a single-mode photon number state of occupation $n=0,1,2,\dots$.
(Here, we assume that all photons have the same frequency $f$, and only one photon polarization couples to the detuning parameter.)

In Eq.~(\ref{Eq:SemiHamiltonian}), the $V$ matrix describes the coupling between the semiclassical driving field and the quantum dot.
For a quantum Hamiltonian, the ac coupling occurs through a single mode of the electric field, whose quantum field operator is given by~\cite{GerryBook, DressedAtom} % DressedAtom P.40
\begin{equation}
\hat{E}_x\propto (a^\dagger +a) . \label{Eq:Vint}
\end{equation}
The $V$ matrix should therefore be replaced by the coupling term $V_\text{int}=V_\text{dot}\otimes(a^\dagger+a)$, where $V_\text{dot}$ is a $3\times 3$ matrix acting on the quantum dot hybrid qubit. 
It is important to note that the characteristic photon state of a semiclassical driving field is not the number state $\ket{n}$, but rather the coherent state $\ket{\alpha}$, defined as the eigenstate of the annihilation operator, $a|\alpha\rangle = e^{-i 2 \pi f\!t} \alpha_0 |\alpha\rangle$, yielding~\cite{DressedAtom} % P. 415, eq B.16.a & p.469, eq (28)
\begin{equation}
|\alpha(t) \rangle = e^{-|\alpha_0|^2/2}\sum_{n=0}^\infty e^{-i 2 \pi n f\!t}\frac{\alpha_0^n}{\sqrt{n!}}\ket{n} .
\label{Eq:CoherentState}
\end{equation}
The average photon occupation of the coherent state, $N$, is given by the expectation value
\begin{equation}
N = \langle \alpha| \hat{N}|\alpha \rangle = \langle \alpha | a^{\dagger} a|\alpha \rangle =  |\alpha_0|^2.  
\label{Eq:N}
\end{equation}
Similarly, we can determine $V_\text{dot}$ from the semiclassical correspondence principle~\cite{DressedAtom}, 
\begin{equation}
V\cos (2\pi f\!t) = \bra{\alpha}V_\text{int}\ket{\alpha}. \label{Eq:Vclassical}
\end{equation}
As discussed in Appendix~\ref{Sec:DressedStateDetail}, this correspondence reduces to $V_\text{dot} = V/ 2 \sqrt{N}$.

\begin{figure}
\includegraphics[width=3in]{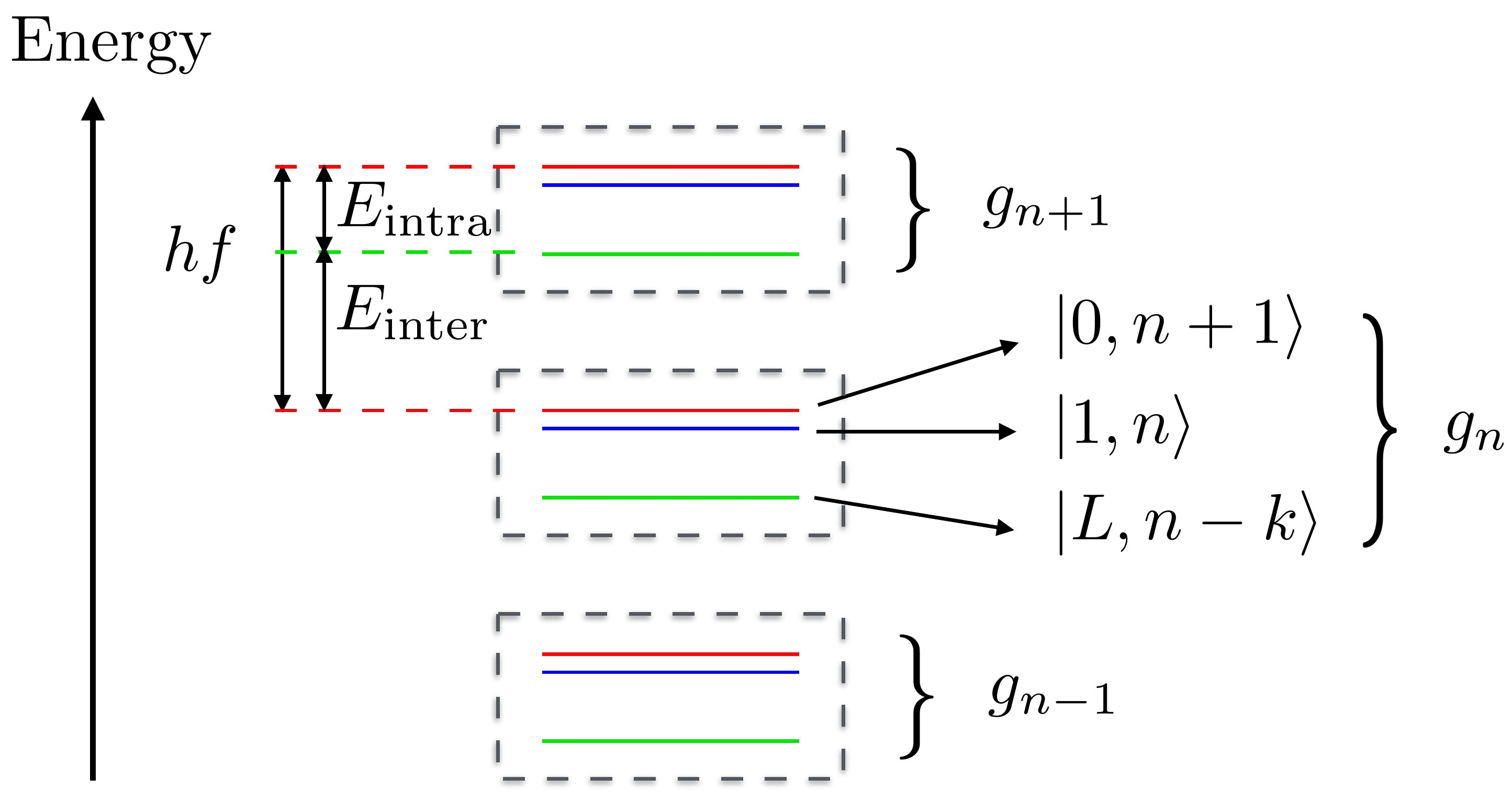}
\caption{
\label{Fig:EnergyLevel}
A cartoon depiction of the manifold structure of bare states of a quantum dot hybrid qubit coupled to a photon field with frequency $f$.
The 3D manifold labeled $g_n$ contains two qubit levels, $\ket{0,n+1}$ and $\ket{1,n}$, which are nearly degenerate near the resonance condition $h\!f \simeq E_1-E_0$, and a leakage level $\ket{L,n-k}$.
Here, $n$ and $k$ indicate photon numbers.
If $E_\text{intra}=h\!f-E_\text{inter}$ represents the energy width of a given manifold, as indicated in the diagram,
then $k$ is defined such that $E_\text{intra}<h\!f/2$. }
\end{figure}

The full quantum Hamiltonian is now given by
\begin{equation}
\Hamiltonian_\text{QM} = \Hamiltonian_\text{dot}+ \Hamiltonian_\text{ph} + V_\text{int} .
\label{Eq:QuantumModel}
\end{equation}
A typical energy spectrum for $\Hamiltonian_\text{dot}+ \Hamiltonian_\text{ph}$  is shown in Fig.~\ref{Fig:EnergyLevel}.
For the case of a quantum dot hybrid qubit, we see that the energy levels divide into manifolds comprised of three levels, which are separated from other manifolds by the energy $h\!f$.
When the qubit is driven near its resonance condition, $ h\!f \simeq E_1-E_0$, the bare states $\ket{0,n+1}$ and $\ket{1,n}$ are nearly degenerate and comprise two of the three levels in the manifold labeled $g_n$.
If $k$ is the value for which $|E_L-E_1 - kh\! f| < h\!f/2 $, then the leakage state $\ket{L,n-k}$ is the third member of $g_n$.
The manifold structure is periodic, and in the absence of interactions ($V_\text{int}=0$), the 3D Hamiltonian for manifold $g_n$ takes the form
\begin{equation}
\Hamiltonian_n= \Hamiltonian_\text{block} + n h\!f,
\label{eq:Hntilde}
\end{equation}
where $\Hamiltonian_\text{block}=\Hamiltonian_0+h\!f(\ket{0}\bra{0}-k\ket{L}\bra{L})$ is independent of $n$.
The interaction term couples bare states that differ by one photon,
\begin{equation}
\langle i, n |V_\text{int}|j, m\rangle = \frac{V_{ij}}{2} (\delta_{n,m+1}+ \delta_{n,m-1}) , \label{eq:Vint}
\end{equation}
yielding hybridized states known as ``dressed" states,
where $V_{ij}$ are elements of the matrix $V$, as defined in Appendix~\ref{sec:LargeDetuning}.

We exploit the manifold structure in Fig.~\ref{Fig:EnergyLevel} by diagonalizing $\Hamiltonian_\text{QM}$ into 3D blocks via a Schrieffer-Wolff unitary transformation~\cite{WinklerBook}.
The full, transformed Hamiltonian, described in Appendix~\ref{Sec:DressedStateDetail}, then reduces to a tensor product of the form
\begin{equation}
\tilde{\Hamiltonian}_\text{QM} = \otimes_n (\tilde\Hamiltonian_\text{block} + nh\!f), \label{Eq:Hblock}
\end{equation}
where $\tilde\Hamiltonian_\text{block}$ is also independent of $n$.
At lowest order in this perturbation theory, we find that $\tilde\Hamiltonian_\text{block}=\Hamiltonian_\text{block} + (V_{01}|0\rangle\langle 1|+h.c.)$ as expected; the leading corrections to $\tilde\Hamiltonian_\text{block}$ occur at order $O[(A/h\!f)^2]$.
However, the block structure of Eq.~(\ref{Eq:Hblock}) extends to all orders, allowing us to obtain simple analytical estimates for the time-evolution operator $\tilde{U}_\text{QM}(t)$.
Finally, we project the solution back onto the 3D qubit subspace via the semiclassical correspondence,
\begin{equation}
U_\text{semi} = \bra{\alpha(t)}\tilde{U}_\text{QM}\ket{\alpha(0)}  . \label{eq:Usemi}
\end{equation}
We note here that $\tilde{U}_\text{QM}$ evolves the full system, including both dot and photon states.

There are several benefits to using a fully quantum Hamiltonian, as we have done here.  
First, the time-varying driving term in the semiclassical Hamiltonian is replaced by a constant coupling to a photon field, allowing us to solve an effectively static Hamiltonian in the bare-state basis set.  
The price we pay for this convenience is a greatly expanded Hilbert space.
The second advantage of using quantized fields is the intuition provided:  the elementary processes of absorption and emission of photons can be readily identified.
  
\subsection{Strong Driving}
The textbook description of spin resonance is obtained in the weak-driving limit~\cite{GerryBook}, where $f_\text{Rabi}\ll f$, and $f_\text{Rabi}\propto A$ is the Rabi frequency.
In that case, $\Hamiltonian_\text{semi}$ can be solved by transforming to the rotating frame and applying the RWA in which we drop the ``counter-rotating" term.
The RWA is equivalent to the lowest-order term of the dressed-state perturbation theory, and its dynamics correspond to smooth circular trajectories on the Bloch sphere.
For quantum dot qubits however, it is often desirable to work in the strong driving regime $f_\text{Rabi}\sim f$ to minimize the effects of decoherence.
In this regime, the qubit dynamics are not smooth; they exhibit additional fast oscillations and other complicated behavior, as discussed below.
The RWA therefore breaks down and the semiclassical approach becomes cumbersome.

Corrections to the RWA can be obtained straightfowardly within the dressed-state formalism by retaining higher-order terms in the perturbation expansion.
These corrections are manifested as renormalizations of (i) the resonance frequency (i.e., the Bloch-Siegert shift~\cite{Bloch1940}), and (ii) the Rabi frequency and gate period.
Such effects are are well known for the case of a simple two-level system with a transverse drive \cite{CohenTannoudji1973, DressedAtom}; in Appendix~\ref{Appendix:two-level}, we reproduce those results using the dressed state methods outlined above.
For a quantum dot hybrid qubit, which is the main focus of this paper, strong driving can also cause additional strong-driving effects, such as occupation of the leakage state, as discussed below.

\begin{figure*}
\includegraphics[width=6in]{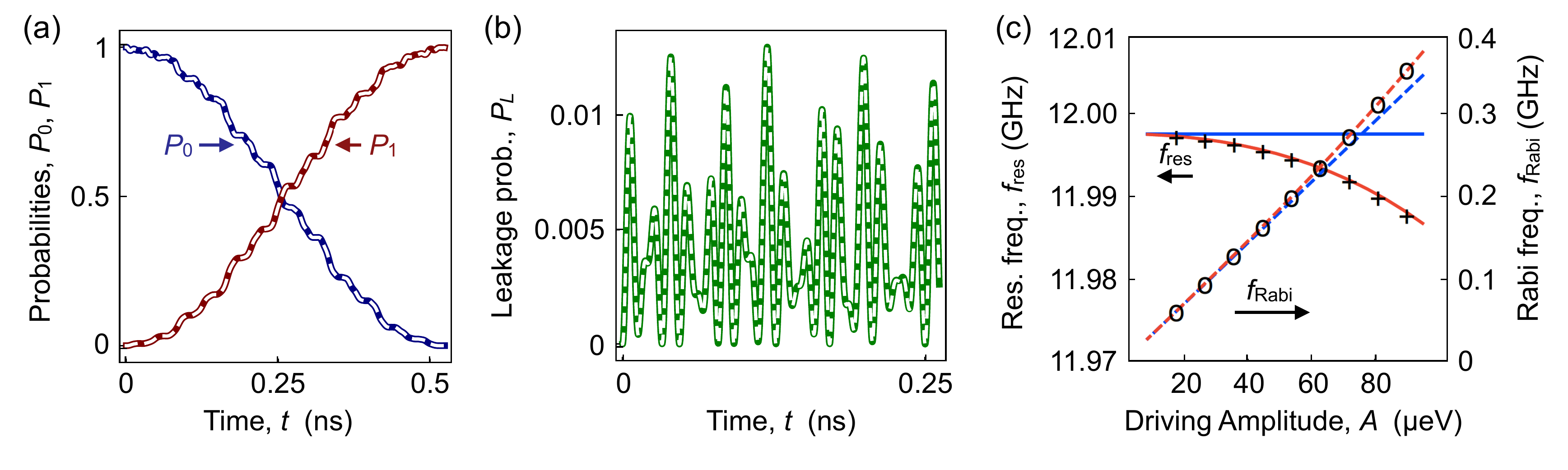}
\caption{
\label{Fig:HybridQubitResult}
Dynamics of a strongly driven quantum dot hybrid qubit.
(a), (b) Numerical solutions of Eq.~(\ref{Eq:SemiHamiltonian}) are plotted as solid lines.
Here, the simulation parameters are given by $E_\text{ST}/h= 12$~GHz, $\{\varepsilon,\Delta_{10}, \Delta_{20}, A\} = \{6 ,0.7 , 0.7, 0.33 \}\times E_{ST}$, and $r_1=r_2=1$, and the ac drive is applied to the tunnel coupling.
The corresponding analytical solutions of the dressed-state theory with terms included up to $O[(A/hf)^3]$ are plotted as dashed white lines.
(a) The probabilities $P_0(t)=|c_0(t)|^2$ and $P_1(t)=|c_1(t)|^2$ of the logical qubit states are plotted for the initial state given by $\ket{0}$. 
The sinusoidal envelopes correspond to the conventional Rabi oscillations that would occur under weak driving (i.e., the RWA), while the fast modulations are caused by strong driving.
The analytical results shown here include corrections to the RWA up to $O[(A/hf)^3]$ and reproduce all observable features in the fast oscillations.
(b) The corresponding evolution of the leakage probability $P_L(t)=|c_L(t)|^2$.
Again, the analytical solutions reproduce all observable features of the numerical simulations, and the square of the difference between the analytical and numerical results is less than $10^{-7}$ over the entire range.
(c) Numerical and analytical calculations of the resonance frequency (analytical results are plotted with solid lines) and Rabi frequency (dashed lines) as a function of driving amplitude, for microwave driving of the detuning, with simulation parameters $E_{ST}/h= 12$~GHz and $\{\varepsilon, \Delta_{1}, \Delta_{2}\} = \{6, 0.7 , 0.7\}\times E_{ST}$.
The simulation results ($+$ markers for the resonance frequency and $\circ$ markers for the Rabi frequency) are obtained by performing a sweep of the driving frequency $f$;
$f_\text{res}$ is identified as the driving frequency that minimizes the Rabi frequency, as in Eq.~(\ref{eq:fRabifres}), while $f_\text{Rabi}$ is identified as the dominant peak in the Fourier spectrum.
Analytical results are shown for the RWA (blue), corresponding to $f_0=(E_1-E_0)/h$, and corrections to the RWA up to $O[(A/h\!f)^4]$ (red).
The difference between the actual resonance frequency and the RWA is called the Bloch-Siegert shift.
Any remaining deviation of the numerical results from the analytical calculations comes from higher-order terms in the perturbation expansion.
}
\end{figure*}

\section{Evolution Of A Quantum Dot Hybrid Qubit Under Strong Driving}\label{sec:results}
We now explore the dynamics of the quantum dot hybrid qubit using two different methods.
First, we perform numerical simulations of the full Hamiltonian given in Eq.~(\ref{Eq:SemiHamiltonian}).
Next, we obtain analytical estimates based on the dressed-state theory, which are derived in Appendix~C and summarized below. 
In both cases, the ac drive is applied to the tunnel coupling $\Delta_\text{ac}$.
We then obtain solutions of the form
\begin{equation}
\ket{\psi (t)} =c_0(t)\ket{0}+c_1(t)\ket{1}+c_L(t)\ket{L} ,
\label{eq:state}
\end{equation}
and compare the results in Figs.~\ref{Fig:HybridQubitResult}(a) and (b).
Detuning driving is also considered in Appendix~\ref{Sec:PulseShapeDetuning}, with some results shown in Fig.~\ref{Fig:HybridQubitResult}(c).

\subsection{Numerical Simulations}
In Figs.~\ref{Fig:HybridQubitResult}(a) and (b) we plot typical oscillations results obtained numerically under strong driving, with a ratio of Rabi to qubit frequencies of about 0.08, for the initial state $c_0(0) = 1$.
The dominant feature observed in Fig.~\ref{Fig:HybridQubitResult}(a) is a slow sinusoidal envelope reflecting the expected Rabi oscillations.
Modulating this smooth behavior, we also observe fast oscillations, which are typical of strong driving.
The fast oscillations have two sources.  
The first is the counter-rotating term in the drive, that are neglected within the RWA.
The second source of oscillations is leakage.
Fig.~\ref{Fig:HybridQubitResult}(b) shows the time-varying occupation of the leakage state, which is directly reflected as reduced occupation of the qubit states.

Strong driving also has other significant effects on the dynamical evolution.
As shown in Fig.~\ref{Fig:HybridQubitResult}(c), these include corrections to the RWA for both the resonant and Rabi frequencies.
In the simulations, we determine the resonant frequency $f_\text{res}$ by minimizing the Rabi frequency $f_\text{Rabi}$ for a fixed driving amplitude $A$.
The resonant and Rabi frequencies are related by the standard relation
\begin{equation}
f_\text{Rabi}=\sqrt{(f-f_\text{res})^2+(V_{01}/h)^2} , \label{eq:fRabifres}
\end{equation}
which is also derived in Eq.~(\ref{Eq:fRabiTLS}).
(As noted in Appendix~\ref{sec:LargeDetuning}, holding $A$ constant is equivalent to holding $V_{01}$ constant here.)
One could alternatively try to identify $f_\text{res}$ by maximizing the oscillation visibility.
Strong driving causes errors in this procedure, however, because the fast oscillating terms also contribute to the visibility at frequencies away from $f_\text{res}$.

\subsection{Analytical Estimates} \label{Sec:HybridQubitResult}

In Appendix~\ref{Sec:HybridQubitDetail}, we derive the time-evolution operator for the quantum dot hybrid qubit using the dressed state method, up to order $O[( A/h\!f)^2]$ in the perturbation expansion.
We demonstrate the accuracy of this approach in Fig.~\ref{Fig:HybridQubitResult}(a) by plotting the analytical results directly on top of the numerical results.
The analytical derivations appear to capture all the fast-oscillating features associated with strong driving.
Even higher accuracy can be achieved by retaining higher orders in the expansion.

The dressed-state theory provides insight into the origins of the fast oscillations, which are caused by couplings between bare states, due to the interaction term, $V_\text{int}$.
For example, leakage is caused by the hybridization of qubit and leakage states, with mixing coefficients of order $O[A/h\!f]$.   
Fast oscillations also arise from the counter-rotating terms in $V_\text{int}$, which hybridize bare states in different $g_n$ manifolds.
The effects of the leakage and counter-rotating terms become prominent in the strong-driving regime, as observed in Figs.~\ref{Fig:HybridQubitResult}(a) and (b), and discussed in Appendix~\ref{Sec:HybridQubitDetail}. 
To reduce control errors in quantum dot hybrid qubits, it is necessary to suppress the fast oscillations via pulse shaping, as described in Sec.~\ref{Sec:PulseShape}.

A secondary effect of the counter-rotating terms has already been noted in Fig.~\ref{Fig:HybridQubitResult}(c), where the resonant frequency differs from the bare qubit energy splitting and becomes a function of the driving amplitude.
This Bloch-Siegert shift arises at order $O[(A/h\!f)^2]$ in the perturbation expansion.
In Appendix~\ref{Sec:HybridQubitDetail} we derive its form as
\begin{multline}
\hspace{-.1in}  h\! \tilde f_\text{res} 
= h\!f_0 
+ \frac{V_{01}^2}{4 (E_1-E_0) } +  \frac{V_{0L}^2}{4 (E_L -E_1 )} - \frac{V_{1L}^2}{4 (E_L -E_0 )}  \\
+  \frac{V_{0L}^2}{4 (E_L +E_1-2E_0)} -  \frac{V_{1L}^2}{4 (E_L +E_0-2E_1 )}  .
\label{Eq:drivingfreq}
\end{multline}
Here, $f_0=(E_1-E_0)/h$ is the bare resonant frequency, consistent with the RWA, and $\tilde f_\text{res}$ is the renormalized frequency, including the Bloch-Siegert shift.
% The terms $V_{ij}$ refer to specific matrix elements of the driving matrix, $V$, with $V_{ij}\propto A$.
Written in terms of the $V_{ij}$, the couplings between states $|0\rangle$, $|1\rangle$, and $|L\rangle$ induced by the driving, Eq.~(\ref{Eq:drivingfreq}) is valid for either detuning or tunnel coupling driving.
Explicit forms for the $V_{ij}$ for detuning and tunnel coupling driving are given in Appendix~\ref{sec:LargeDetuning}. 
In Fig.~\ref{Fig:HybridQubitResult}(c), we plot our analytical estimates for $\tilde f_\text{res}$ for the case of detuning driving, keeping terms up to order $O[(A/h\!f)^4]$.
The dressed-state theory describes the resonant frequency shifts with very high accuracy.

It is interesting to compare the Bloch-Siegert shift of the quantum dot hybrid qubit, in Eq.~(\ref{Eq:drivingfreq}), with that of a simple, transversely driven two-level system.
The latter is derived in Appendix~\ref{Sec:DressedStateDetail} and Ref.~\cite{DressedAtom}, giving
\begin{equation}
h\! \tilde f_\text{res} = h\!f_0 + \frac{V_{01}^2}{4(E_1-E_0)}. \label{Eq:f1twolevel}
\end{equation}
In this case, the only correction to the RWA comes from the counter-rotating term.
The additional terms in Eq.~(\ref{Eq:drivingfreq}) are therefore caused by leakage, as apparent from their functional forms.
For both types of corrections, the energy shift amounts to a dynamical repulsion between the qubit energy levels, which grows with the driving amplitude.

The Rabi frequency is also renormalized under strong driving.
The leading-order expression for the Rabi frequency, Eq.~(\ref{eq:fRabifres}), is consistent with the RWA, and reduces to $h\!f_\text{Rabi} = V_{01}$ at resonance.
We can go beyond this level of approximation to obtain corrections to $f_\text{Rabi}$ at $O[(A/h\!f)^3]$, which are plotted in Fig.~\ref{Fig:HybridQubitResult}(c), and which account for the small splitting between our numerical results and the RWA prediction.
The derivation of these higher order corrections is tedious but straightforward, and is not reported here.

Fast oscillations due to strong driving are typically not taken into account when implementing gate operations, resulting in potential control errors. 
In principle, these errors could be suppressed to arbitrary order by including appropriate corrections to the resonant frequency and accounting for the more complicated dynamics shown in Fig.~\ref{Fig:HybridQubitResult}. 
In practice this is impractical, particularly since the fast oscillation frequency is rather high ($\gtrsim 10$~GHz).
We can estimate the control errors that would occur if the Rabi oscillations were assumed to be smooth and sinusoidal, as in the RWA.
If we explicitly consider an $X_\pi$ gate acting on the initial state $c_0(0)=1$, with gate period $t_g$, then the ideal final state (without control errors) would be $c_0(t_g)=0$.
Any deviation of $c_0(t_g)$ from zero therefore characterizes the control error.
(Note that this represents a state fidelity calculation; in Sec.~\ref{Sec:fidelity}, we compute the full process fidelity for such an $X_\pi$ gate.)
For simplicity here, we consider the far-detuned regime, $\varepsilon\gg E_\text{ST}$, as is typical for experiments.
Using the result of Eq.~(\ref{eq:P0hybrid}), we find that the control error for this process therefore scales as $|c_0(t_g)|^2\sim (A\Delta/\varepsilon E_\text{ST})^2$ for tunnel coupling driving, or $(A\Delta^2/\varepsilon^2E_\text{ST})^2$ for detuning driving, where $\Delta=\Delta_{1,2}$ represents a typical tunnel coupling.

These scaling estimates suggest that control errors due to strong driving could potentially be suppressed by reducing the driving amplitude $A$.
Unfortunately, this is not possible when gate times $t_g$ are held constant to avoid errors caused by decoherence.
To see this, we note from Appendix~\ref{Sec:HybridQubitDetail} that $t_g\propto\varepsilon/A$ for tunnel coupling driving, or $\varepsilon^2/A$ for detuning driving, so $|c_0(t_g)|^2\propto t_g^{-2}$ in both cases.
Hence, if $t_g$ is held constant, it is impossible to independently suppress $|c_0(t_g)|^2$.

To summarize this section, a dressed-state formalism may be used to enhance quantum gate fidelities by providing corrections to the resonant and Rabi frequencies.
However, fast oscillations cannot be avoided by the gating schemes discussed so far, causing potential control errors.
In the following section, we show that pulse shaping can ameliorate this problem.

\section{Suppressing Fast Oscillations via Pulse Shaping \label{Sec:PulseShape}}
In Sec.~\ref{sec:results}, we showed that strong driving induces fast oscillations, making it difficult to control qubit gate operations.
We also showed that it is impossible to suppress fast oscillations by simply reducing the driving amplitude while holding $t_g$ fixed.
Here, we show that simple pulse shapes can improve the gate fidelity significantly.
In particular, we consider a scheme where $t_g$ is held fixed, but the driving amplitude is turned on and off smoothly, at the beginning and end of the gate pulse.  

The benefits of using continuous, nonsingular pulse shapes are twofold.
First, as shown in Fig.~\ref{Fig:PulseShape}, singular pulses generate Fourier spectra with broad peaks and increased weight at high frequencies, causing unwanted leakage.
[For simplicity, no ac drive was included in Fig.~\ref{Fig:PulseShape}(b); when driven, the central peak splits into two peaks, centered at the frequencies $\pm f$.]
The specific shape of the pulse determines the spectral density at high frequencies, but both of the continuous pulses in the figure exhibit significantly lower density at high frequencies than the rectangular pulse, which was implicitly assumed in Sec.~\ref{sec:results}.
The second benefit of a continuous pulse is that it suppresses fast oscillations at the beginning and end of a gate (near $t=0$ and $t_g$), where the control errors occur.  
This is because the amplitude of the fast oscillations is proportional to the pulse envelope $A(t)$.
\red{
As $A(t)$ goes to zero near the end points of the pulse, the amplitude of the fast oscillations also vanishes.}
Below, we show that these simple modifications of the pulse shape yield significant improvements in the gate fidelity.

\subsection{Pulse Shapes}\label{Sec:shapes}
\begin{figure}
\includegraphics[width=3in]{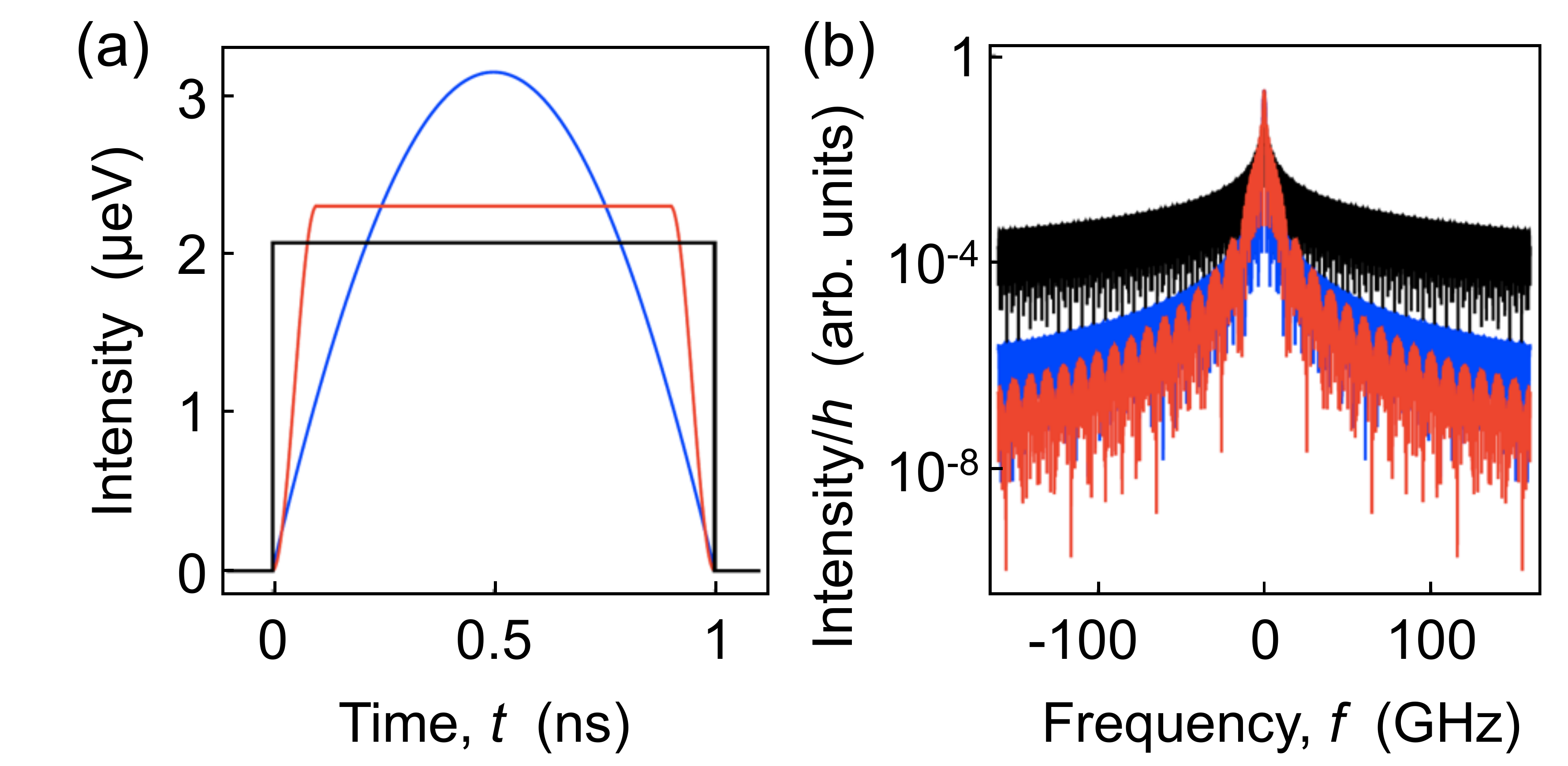}
\caption{
\label{Fig:PulseShape}
Pulse envelopes and their Fourier spectra.
(a) The three pulse shapes considered in this work are  rectangular (black), truncated Gaussian (blue), and smoothed rectangular (red), as defined in Eqs.~(\ref{Eq:Rectangular})-(\ref{Eq:SmoothRectangular}).
The parameters used to generate the pulses shown here are $\{t_g, \sigma, t_r\} = \{1,1,0.1\}$ ns, where $t_g$ is the gate time, $\sigma$ sets the width of the Gaussian pulse, and $t_r$ is the rise time of the smoothed rectangular pulse.
(b) The Fourier spectra of the same three envelopes, using the same color scheme.
Here, the frequencies $f \simeq (E_L-E_0)/h \pm (E_1-E_0)/h$, and $f\simeq (E_L-E_1)/h\pm (E_1-E_0)/h$ are associate with leakage; in our simulation, these are given by $48$, $60$, $72$, and $84$~GHz. 
At such high frequencies, the smoothed rectangular pulse has the lowest spectral density, and should therefore be the most effective at suppressing leakage. }
\end{figure}

% In this work, the ac drive can be applied to either the detuning or the tunnel coupling.
In our simulations, we consider several different pulse envelopes $A(t)$.
Here, $A(t)$ corresponds to one of the experimentally tunable parameters, such that $A(t) \cos (2 \pi f\!t)=\{ \varepsilon(t)\text{ or }\Delta_i(t) \}$. 
Because the Rabi frequency is determined by $V_{01}$, as specified in Appendix~\ref{sec:LargeDetuning}, and since $V_{01}\propto A$, it is convenient to treat $V_{01}(t)$ as the tunable parameter in the following discussion.
When performed on resonance, each of the pulse shapes yields a rotation about the $\hat{x}$ axis.
The total angle of rotation $\theta$ is approximately given by the relation $\theta=\int_0^{t_g}(V_{01}/\hbar)dt$. 
Here, we specifically consider $X_\pi$ rotations, with pulses normalized to have the same gate  time $t_g$.  
(We also consider $X_{\pi/2}$ rotations in Appendix~\ref{Sec:XPI2}, obtaining qualitatively similar results.)
We then compare the pulse shapes by computing their gate fidelities as a function of $t_g$.
Note that for continuous pulse envelopes, the Hamiltonian at different times does not commute with itself, so the relation between $t_g$ and $\theta$ given above is inexact.
In our simulations, however, it is a very good approximation, yielding gates with high fidelities.
The three pulse shapes shown in Fig.~\ref{Fig:PulseShape} are defined as follows.

\begin{figure*}
\includegraphics[width=6in]{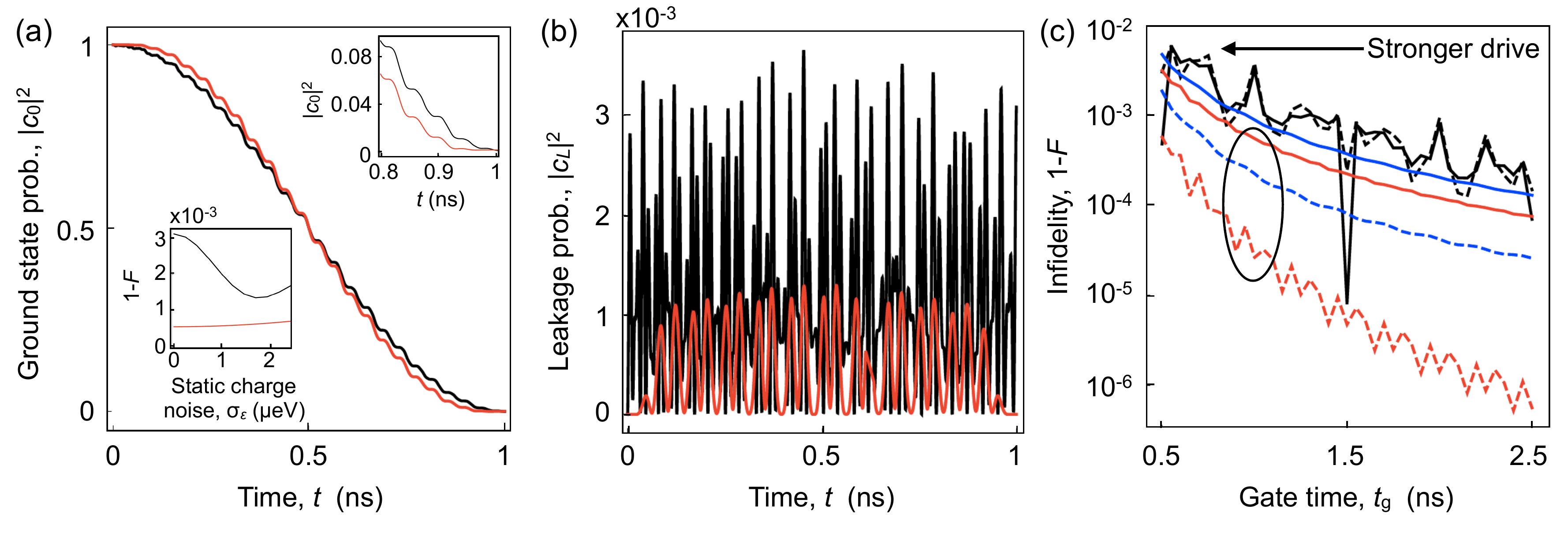}
\caption{
\label{Fig:Results}
Improving the fidelity of $X_\pi$ rotations using pulse shaping in the strong-driving regime.
Simulations are performed by driving the tunnel coupling with the control parameters $E_\text{ST}/h = 12$ GHz, $\{\varepsilon,\Delta_{10},\Delta_{20}\} = \{6,0.7,0.7\} E_\text{ST}$, and $r_1=r_2=1$. Here, $E_\text{ST}$ is the singlet-triplet energy splitting of the doubly occupied dot, $\varepsilon$ is the detuning between the dots, $\Delta_{10}$ ($\Delta_{20}$) are the time-independent part of the tunnel couplings between the states $\ket{S\cdot}$ and $\ket{\cdot S}$ ($\ket{\cdot T}$), and $r_1$ ($r_2$) are the coefficients for the responses of the tunnel couplings to the AC signal.
(a) and (b) show typical dynamical evolutions for an $X_\pi$ gate with gate time $t_g = 1$~ns, for the initial state with state amplitudes (see Eq.~(\ref{eq:state})) of $c_0=1$, $c_1=c_L=0$.
Two results are shown, corresponding to a sharp rectangular pulse (black) or a smoothed rectangular pulse (red).
The upper inset  of (a) shows a blow up of times near $t=t_g$, where the smoothed pulse suppresses the fast oscillations.
The lower inset shows estimates of $1-F$, where $F$ is the fidelity, for simulations including quasistatic charge noise of uniform distribution with zero mean and standard deviation $\sigma_{\varepsilon}$.
In (b), we see that the leakage state occupation is suppressed over the whole gate evolution for the smoothed pulse, particularly near
$t=0,t_g$.
\red{This is because leakage probability depends quadratically on the pulse envelope, $|c_L|^2 \sim A^2(t)$, and therefore vanishes when $A(t)\rightarrow 0$ near the endpoints of the pulse. }
(c) shows the process infidelity, $1-F$,~\cite{ChuangBook} computed for several different scenarios:
rectangular (black), truncated Gaussian (blue), and smoothed rectangular (red) pulse shapes with no dynamical corrections (RWA, solid lines), and dynamical corrections for $\tilde f_\text{res}$ up to order $O[(A/h\!f)^2]$ (dashed lines). 
Using these smooth pulse shapes and strong-driving corrections, we can achieve gate fidelities $>99.9\%$ for several different scenarios, with gate times as short as 1~ns (black oval).  
For the smoothed rectangular pulse, the fidelity is $>99.99\%$.  (Note that these results, except the lower inset of (a), do not include noise.)
}  
\end{figure*}
 
1.  The \emph{rectangular pulse} is defined as
\begin{equation}
V_{01} = h/2t_g,
\label{Eq:Rectangular} 
\end{equation}
when $0 \leq t \leq t_g$, and zero otherwise. 
Since $V_{01}(t)$ is piecewise constant here, we are able to apply the dressed-state formalism to obtain analytic corrections to the resonant and Rabi frequencies, as discussed in Sec.~\ref{Sec:HybridQubitResult}.

2.  A \emph{truncated Gaussian pulse} has recently been employed for leakage suppression~\cite{Gambetta2011}.
Its form is given by
\begin{equation}
V_{01} = \frac{h}{2} \frac{\exp [-(t-t_g/2)^2/2 \sigma^2]-\exp [-t_g^2/8 \sigma^2]}{\sqrt{2 \pi \sigma^2} \erf [t_g/\sqrt{8}\sigma] -t_g \exp[-t_g^2 / 8 \sigma^2]},
\label{Eq:Gaussian}
\end{equation}
when $0 \leq t \leq t_g$, and zero otherwise.
The pulse has a characteristic width of $2\sigma$ when $\sigma\ll t_g$, and it has no discontinuities.
An example is shown in Fig.~\ref{Fig:PulseShape}.
Since $V_{01}$ is continuous in time, its high-frequency spectrum has a lower density than the rectangular pulse.
However, since $dV_{01}/dt$ is discontinuous at the endpoints of the pulse, we expect to observe more spectral weight at high frequencies than for a pulse with a continuous second derivative.
(This discontinuity is suppressed when $\sigma\ll t_g$.)

3.  A \emph{``smoothed"} \emph{rectangular pulse} is obtained by replacing the singular steps with sinusoids \cite{PhysRevLett.115.133601,PhysRevA.94.032323}.
In this case,
\begin{equation}
V_{01} = \left\{ 
\begin{array}{cc} \vspace{0.03in}
\frac{h[1- \cos(\pi t/t_r) ]}{4(t_g-t_r)} &  (0\leq t \leq t_r), \\ \vspace{0.03in}
\frac{h}{2(t_g-t_r)} & ( t_r< t < t_g - t_r), \\
\frac{h[1+ \cos(\pi [t-t_g+t_r]/t_r) ]}{ 4(t_g-t_r)} \hspace{0.2in} &  (t_g-t_r \leq t \leq t_g), 
\end{array}
\right.
\label{Eq:SmoothRectangular}
\end{equation}
and zero otherwise.
Here, $t_r$ is the ramp time, and we assume that $2t_r<t_g$.
An example is shown in Fig.~\ref{Fig:PulseShape}.
Since this pulse is completely smooth, it has less spectral weight at high frequencies than either of the previous shapes.
Moreover, since the pulse is nearly rectangular, the renormalized resonant and Rabi frequencies obtained from the dressed-state theory should be accurate over most of the gate duration.

\subsection{Simulations of Gate Fidelities}\label{Sec:fidelity}
In this section, we compute the process fidelity for quantum gates obtained using the pulse shapes shown in Fig.~\ref{Fig:PulseShape}.
We consider scenarios with or without the dressed-state corrections for strong driving.
Our results for $X_\pi$ rotations based on tunnel coupling driving are plotted in Fig.~\ref{Fig:Results}.
Additional results for detuning driving and $X_{\pi/2}$ rotations are reported in Appendices~\ref{Sec:PulseShapeDetuning} and \ref{Sec:XPI2}, respectively.

We first perform dynamical gate simulations by solving the time-dependent Schr\"odinger equation $\dot{\rho}(t)=-(i/\hbar)[\Hamiltonian_\text{semi},\rho]$.
Here $\rho$ is a $3\times3$ density matrix describing both the logical and leakage states.
Following Ref.~\cite{ChuangBook}, if $\rho_0$ represents the initial density matrix before a gate operation, and $\mathcal{E}$ represents the final density matrix after the gate operation, then the initial and final density matrices can be related by the process matrix $\chi$ via
\begin{equation}
\mathcal{E} = \sum\limits_{m,n} E_m \rho_0 E_n ^{\dagger} \chi_{mn},
\end{equation}
where $\{E_m\}$ is a basis for the vector space of $3\times 3$ matrices.
The process fidelity is then defined as $F=\text{Tr} [\chi_\text{sim} \chi_\text{ideal}]$, where $\chi_\text{sim}$ is the actual process matrix for the simulations, including strong-driving effects, and $\chi_\text{ideal}$ describes the ideal rotation.
Since $\chi_\text{ideal}$ does not involve the leakage channel, it is easy to show that $\text{Tr} [\chi_\text{sim} \chi_\text{ideal}]$ also does not contain any information about leakage processes in $\chi_\text{sim}$; to compute $F$, it is therefore sufficient to project $\rho_0$ and $\mathcal{E}$ onto the 2D logical subspace and solve for $\chi$ matrices that are $4\times 4$.  
In this case, we choose $E_m$ from the Pauli basis $\{I, \sigma_x, -i \sigma_y, \sigma_z\}$ and follow the standard procedure for computing $F$~\cite{ChuangBook}.

Typical simulation results for an $X_\pi$ gate are shown in Figs.~\ref{Fig:Results}(a) and (b) as a function of time $t$, between $t=0$ and the final gate time, given by $t_g=1$~ns in these simulations.
The initial state is given by Eq.~(\ref{eq:state}) with  $c_0(0)=1$, $c_1(0)=c_L(0)=0$.
Two simulation results are shown.
The first (black curve) uses a conventional rectangular pulse shape, while the second (red curve) assumes identical parameters for a smoothed rectangular pulse.
The key difference between the two evolutions can be seen in the upper inset of Fig.~\ref{Fig:Results}(a), where the fast oscillations of the smoothed pulse are strongly suppressed at times $t=0$ and $t_g$, compared to the rectangular pulse.
In Fig.~\ref{Fig:Results}(b), we see that the smooth pulse suppresses leakage oscillations over the entire gate period, but especially \red{near the endpoints}. 
\red{As explained in Appendix~\ref{Sec:HybridQubitDetail}, this is because the leakage probability depends quadratically on the pulse envelope, $|c_L|^2 \sim A^2(t)$. 
As $A(t)$ approaches zero near its endpoints, the amplitude of the leakage oscillations also vanishes. 
This perfect cancellation is a consequence of noise-free evolution, since leakage is then fully coherent.
When noise is present, the cancellation effect is imperfect, and the leakage state becomes slowly occupied over time, even when using a continuous pulse shape; such behavior is outside the scope of the present analysis, however.
} 

To compute the fidelity $F$, the process matrices $\chi_\text{sim}$ and $\chi_\text{ideal}$ should both be expressed in the same reference frame.
Here, $\chi_\text{sim}$ is computed in the lab frame, while $\chi_\text{ideal}$ is defined in the frame rotating at the driving frequency.
The latter must therefore be transformed back to the lab frame.
However, the driving frequency $f$ is not necessarily resonant, depending on the approximations used to calculate $\tilde f_\text{res}$, and this must be incorporated into our definition of $\chi_\text{ideal}$.
For example, if we consider the ideal rotation $X_\pi=-i\sigma_x$ in the rotating frame, the corresponding transformation in the lab frame is given by
\begin{equation}
U_\text{ideal} = 
\begin{pmatrix}
0 & - i e^{ - i \tilde{E}_0 t_g/\hbar}\\
-i e^{ - i\tilde{E}_1 t_g/\hbar} & 0
\end{pmatrix} 
\end{equation}
where $\tilde{E}_0$ and $\tilde{E}_1$ are dynamically renormalized qubit energies in some approximation scheme.
In our simulations, we adopt two different approximations for $\tilde{E}_0$ and $\tilde{E}_1$.
First, we consider the RWA, for which $\tilde{E}_0=E_0$, $\tilde{E}_1=E_1$, and $\tilde f_\text{res}=E_1-E_0$.
Alternatively, we include the dressed-state corrections, defined as $\tilde{E}_0=E_0+\beta_0$, $\tilde{E}_1=E_1+\beta_1$, and $\tilde f_\text{res}=E_1-E_0+\beta_1-\beta_0$, where $\beta_0$ and $\beta_1$ represent the Bloch-Siegert shifts (see Appendix~\ref{Sec:HybridQubitDetail}).

The resulting process infidelities for $X_\pi$ gates are shown in Fig.~\ref{Fig:Results}(c) as a function of the gate time.
(Note that $\text{infidelity}=1-\text{fidelity}$.)
Here we compare the effectiveness of the various scenarios considered in this work.
First, note that the downward trend of the curves is explained by the fact that shorter gate times require stronger driving, which results in worse fidelities.
Second, by comparing the results for rectangular pulses (the two black curves), we see that those fidelities are not particularly improved by including Bloch-Siegert corrections to the resonance frequency, despite the fact that the corrections were derived specifically for rectangular pulses.
This indicates that control errors caused by fast oscillations and leakage are the dominant sources of error for this pulse shape.
This is confirmed by comparing the two other pulse shapes, which generally exhibit higher fidelities, even without including dynamical corrections.

Comparing the truncated Gaussian and smoothed rectangular pulse shapes, we see that the latter yields slightly better fidelities in the absence of strong-driving corrections.
When Bloch-Siegert corrections are included, however, the smoothed rectangular pulse yields significantly better results, reflecting the fact that the dynamical corrections were derived specifically for rectangular pulses.
The Gaussian pulse fidelity also improves when we include frequency corrections.
Comparing all these results, we find that the smoothed rectangular pulse with the renormalized driving frequency yields the best fidelity, with $1-F\lesssim 10^{-4}$ for a 1~ns gate. 

The simulations in Fig.~\ref{Fig:Results} correspond to $X_\pi$ rotations with the ac drive applied to the tunnel coupling.  
To show that similar results hold for other gate conditions, we have performed additional simulations, which we now summarize.
First, we consider gates with the ac drive applied to the detuning parameter, as described in Appendix~\ref{Sec:PulseShapeDetuning} and Fig.~\ref{Fig:HybridQubitResult}(c). 
In this case, we find that the fidelities are generally worse than for tunnel coupling driving.
To understand this, we recall our previous estimate that gate times should scale as $t_g\propto\varepsilon/A$ for tunnel coupling driving, or $\varepsilon^2/A$ for detuning driving. 
In the latter case, holding $t_g$ fixed in the large-detuning limit requires a much larger driving amplitude $A$, yielding lower gate fidelities due to strong driving effects.
Second, we consider $X_{\pi/2}$ rotations for both tunnel coupling and detuning driving, as described in Appendix~\ref{Sec:XPI2}.
The resulting fidelities are slightly better than for $X_{\pi}$ rotations.
This is also easy to understand, because for fixed gate times, an $X_{\pi/2}$ gate requires approximately half the driving amplitude of an $X_\pi$ gate, yielding a higher gate fidelity.

In practice, gate errors depend on an interplay between fast oscillations and environmental noise.
As noted in Sec.~\ref{Sec:HybridQubitResult}, in the absence of noise, strong driving effects could potentially be ameliorated through a detailed knowledge of the evolution, but detuning shifts from environmental noise will change the gate speed and positions of oscillation peaks, so high fidelity can be achieved reliably only if the amplitude of the fast oscillations is suppressed.
To characterize this effect and the ability of shaped pulses to suppress it, we have performed simulations that include quasistatic noise in the detuning parameter.
The lower inset of Fig.~\ref{Fig:Results}(a) shows the results of such simulations for the same parameters as the main panel.
Here we plot the gate infidelity as a function of the standard deviation $\sigma_{\varepsilon}$ of a uniformly distributed detuning noise with zero mean. 
For low noise levels, the smoothed rectangular pulse suppresses leakage errors, as consistent with our previous discussion.
As the noise increases, the smooth pulse shape is still able to suppress errors caused by fast oscillations.
Interestingly, the gate fidelity for sharp rectangular pulses seems to improve with noise.
We attribute this to a beating effect caused by the fast oscillations.
We note that high-frequency noise can also harm qubit coherence under ac driving~\cite{Yan2013}; however, we do not explore this problem here.

\section{Conclusions\label{Sec:conclusion}}
The need for fast gates in quantum dot qubits, including quantum dot hybrid qubits, necessitates the use of strong driving.
We have shown here that the fast oscillations can be fully understood using a dressed-state theoretical formulism.
In principle, these fast oscillations could present a challenge for accurate control, resulting in gating errors.
However, we have shown that fast oscillations, as well as leakage, can be largely suppressed by shaping the pulse envelopes.  
To lowest order, the key to successful pulse shaping is not the precise shapes of the envelopes, but rather their smooth features, which suggests that they could be very simple to implement experimentally.

The most important effect of strong driving on gate fidelities is the dynamical shift of the resonance frequency caused by the counter-rotating term.
In experiments, this Bloch-Siegert shift can be characterized empirically by sweeping the driving frequency at fixed microwave power and identifying the minimum Rabi frequency.
Here, we have used the same empirical method to analyze our simulations.
We have also predicted the Bloch-Siegert shift analytically by applying a dressed-state perturbation theory.
We have used the latter approach here to analyze the unitary evolution of a quantum dot hybrid qubit and estimate the upper bound on the gate fidelity for $X$ rotations.
By performing simulations that include pulse shaping but no decoherence, we predict that fast, high-fidelity gates should be attainable under strong driving, with gate times less than 1~ns, and gate errors below 0.01\%.
Moreoever, we predict that applying the microwave drive to the tunnel coupling rather than the detuning should improve the gate fidelity, since the latter requires a larger driving amplitude to achieve the same gate speed in the large-detuning regime.
For the decoherence rates observed in recent experiments~\cite{KimWardSimmonsEtAl2015}, we therefore expect that environmental noise, not gating errors, should remain the dominant challenge for quantum dot hybrid qubits in the foreseeable future.

Finally, we point out that the dressed-state theory was developed here in the context of quantum dot hybrid qubits.
However, we have also presented the formalism  in a more general form in Appendix~\ref{Sec:DressedStateDetail}, so that it may be applied to other physical systems 
\cite{Fuchs1520,Barfuss2015,Oliver2005,PhysRevB.94.161302}.
For example, in Appendix~\ref{Appendix:two-level} we obtain results for the case of a simple two-level system.

\section{Acknowledgments}
We thank F.~Wilhelm and M. Eriksson for helpful discussions.
Y.-C. Y. was supported by a Jeff and Lily Chen Distinguished Graduate Fellowship. 
The research was also supported by ARO under award no.\ W911NF-12-0607, by NSF under award no.\ PHY-1104660, and by the
Vannevar Bush Faculty Fellowship program sponsored by the Basic Research Office of the Assistant Secretary of Defense for Research and Engineering and funded by the Office of Naval Research through grant no.\ N00014-15-1-0029.

\appendix
\section{Effective Two-level Hamiltonian\label{Sec:EffectiveTwoLevel}}
In this Appendix, we derive an effective 2D Hamiltonian to describe the logical states of the quantum dot hybrid qubit, starting from the full 3D Hamiltonian in Eq.~(\ref{Eq:Hamiltonian}). 
The approximations are accurate over the entire range of detuning values and provide a useful starting point for analyzing adiabatic energy splittings and dc pulsed gates.  However, the reduced Hamiltonian cannot be used to describe ac resonant gates in the strong driving regime, as discussed below.

We begin with the 3D quantum dot hybrid qubit Hamiltonian expressed in the basis set $\{|\cdot S\rangle, |\cdot T\rangle, |S\,\cdot \rangle\}$, as given  in Eq.~(\ref{Eq:Hamiltonian}).
The transformation proceeds in two steps. 
First, we consider the limit $\Delta_1 \rightarrow 0$, with no restrictions on $\Delta_2$.
Hamiltonian~(\ref{Eq:Hamiltonian}) then diagonalizes into two blocks.
It can be further diagonalized into eigenstates $\{\ket{a}, \ket{b}, \ket{c}\}$ via the unitary transformation
\begin{equation}
U_\text{d} =
\begin{pmatrix}
1 & 0 & 0 \\
0 & \frac{\sqrt{E_{L} - E_\text{ST} + \varepsilon}}{\sqrt{2 E_{L}}} & \frac{\sqrt{E_{L} + E_\text{ST} - \varepsilon}}{\sqrt{2 E_{L}}} \\
0 & \frac{\sqrt{E_{L} + E_\text{ST} - \varepsilon}}{\sqrt{2 E_{L}}} & -\frac{\sqrt{E_{L} - E_\text{ST} + \varepsilon}}{\sqrt{2 E_{L}}} 
\end{pmatrix},
\end{equation}
where
\begin{equation}
E_{L} = \sqrt{4 \Delta_2^2 + (E_\text{ST} - \varepsilon)^2}
\end{equation}
is the energy splitting between $\ket{b}$ and $\ket{c}$.
Expressing the full Hamiltonian, with $\Delta_1\neq 0$, in the $\{\ket{a}, \ket{b}, \ket{c}\}$ basis yields
\begin{equation}
\Hamiltonian \!\! = \!\!
\begin{pmatrix}
- \frac{\varepsilon}{2} & \frac{\Delta_1\sqrt{E_{L}+E_\text{ST}-\varepsilon}}{\sqrt{2 E_{L}}}
& -\frac{\Delta_1 \sqrt{E_{L}-E_\text{ST}+\varepsilon}}{\sqrt{2 E_{L}}} \\
\frac{\Delta_1\sqrt{E_{L}+E_\text{ST}-\varepsilon}}{\sqrt{2 E_{L}}} & \frac{E_\text{ST}-E_{L}}{2} & 0 \\
-\frac{\Delta_1\sqrt{E_{L}-E_\text{ST}+\varepsilon}}{\sqrt{2 E_{L}}} & 0 & \frac{E_\text{ST}+E_{L}}{2} 
\end{pmatrix}.
\label{eq:Hfirststep}
\end{equation}

In the second step, we apply a Schrieffer-Wolff transformation $U_\text{SW}$ to approximately block-diagonalize Eq.~(\ref{eq:Hfirststep}) into its logical and leakage subspaces~\cite{WinklerBook}.  Here the logical space corresponds to the lowest two states in Fig.~\ref{Fig:HybridQubit}(c).
To leading order, the new logical basis is given by
\begin{widetext}
\begin{gather}
\ket{\tilde a} = \left(1-\frac{\Delta_{1}^2 (E_L -E_\text{ST} + \varepsilon)}{E_L (E_L +E_\text{ST} + \varepsilon)^2} \right) |\cdot S\rangle 
 + \frac{2 \Delta_1 \Delta_2}{E_{L} (E_{L} + E_\text{ST} + \varepsilon)} |\cdot T\rangle 
- \frac{\Delta_1 (E_{L} - E_\text{ST} + \varepsilon) }{E_{L} (E_{L} + E_\text{ST} + \varepsilon)} |S\,\cdot\rangle,\\
\ket{\tilde b} = 
\left(\frac{\sqrt{E_{L}-E_\text{ST}+\varepsilon}}{\sqrt{2 E_{L}}} + \frac{\sqrt{2} \Delta_{1}^2 \Delta_{2}\sqrt{E_{L}+E_\text{ST}-\varepsilon}}{E_{L}^{5/2}(E_L + E_\text{ST} + \varepsilon )} \right)|\cdot T\rangle  
+ \left( \frac{\sqrt{E_{L}+E_\text{ST}-\varepsilon}}{\sqrt{2 E_{L}}} - \frac{\sqrt{2} \Delta_{1}^2 \Delta_{2}\sqrt{E_{L}-E_\text{ST}+\varepsilon}}{E_{L}^{5/2}(E_L + E_\text{ST} + \varepsilon )}\right)|S\,\cdot\rangle ,
\end{gather}
\end{widetext}
and the effective 2D Hamiltonian in this basis is given by
\begin{equation}
\Hamiltonian_\text{eff} \simeq 
\begin{pmatrix}
-\frac{\varepsilon}{2} - \frac{\Delta_1^2 (E_{L} - E_\text{ST} + \varepsilon)}{E_{L} (E_\text{L} + E_\text{ST} + \varepsilon)}
& \frac{\Delta_1 \sqrt{E_{L} + E_\text{ST} -\varepsilon}}{\sqrt{2 E_{L}}} \\
\frac{\Delta_1 \sqrt{E_{L} + E_\text{ST} -\varepsilon}}{\sqrt{2 E_{L}}} 
& \frac{E_\text{ST} - E_{L}}{2}\\
\end{pmatrix}.
\label{Eq:Heff}
\end{equation}

Equation~(\ref{Eq:Heff}) can be diagonalized to provide a faithful representation of the static energy levels of the logical states, as shown in Fig.~\ref{Fig:HybridQubit}(c). 
Moreover, the Schrieffer-Wolff transformation can be performed to higher orders to achieve even better accuracy.
It is therefore tempting to replace Eq.~(\ref{Eq:Hamiltonian}) by (\ref{Eq:Heff}) in the remainder of our analysis.
However, this procedure is only appropriate for a time-independent Hamiltonian.
In contrast, the transformation matrices $U_\text{d}$ and $U_\text{SW}$ used to the derive $\Hamiltonian_\text{eff}$ are themselves functions of the driving parameters $\varepsilon$, $\Delta_1$, and $\Delta_2$, and are therefore time-dependent.
In this case, the full transformation is given by $U=U_\text{SW}U_d$, and the time-dependent Hamiltonian in the transformed frame is given by
\begin{equation}
\Hamiltonian_\text{eff} =  U^{\dagger}\, \Hamiltonian U - i \hbar U^{\dagger}  \frac{d}{dt} U .
\end{equation}
The final term in this equation is directly proportional to the driving amplitude, and cannot be neglected in the strong-driving regime.
Moreover, this driving term include coupling between the logical and leakage states.
As a result, the 2D description of the dynamics of $\Hamiltonian_\text{eff}$ is necessarily incomplete, and applicable only in the weak-driving regime.
To move beyond this approach, we develop a dressed-state theory in Appendix~\ref{Sec:DressedStateDetail}, extending the full 3D Hilbert space to include microwave photons. 
The formalism provides a means for including strong-driving effects perturbatively and consistently, as discussed in the main text.

\section{Dressed-State Formalism \label{Sec:DressedStateDetail}}
In this Appendix, we provide details on our dressed-state approach for solving the time evolution of a driven qubit.
We first outline the formalism.
We then apply the formalism to a simple example:  a transversely driven two-level system.
We note that similar calculation can also be performed using Floquet theory~\cite{doi:10.1063/1.445483,0022-3700-17-10-015,doi:10.1021/j150665a002,Shirley1965}.

\subsection{Solution Procedure}\label{sec:procedure}
The dressed-state method is described briefly in the main text.  
For completeness, we summarize the solution procedure here.
\begin{enumerate}[label=\emph{\alph*}.]
	\item Diagonalize the semiclassical Hamiltonian with no driving term, yielding the adiabatic eigenbasis $\{\ket{i}\}=\{\ket{0},\ket{1},\ket{L},\dots\}$, comprised of the two logical states, and all other accessible leakage states.
	The resulting diagonal Hamiltonian is defined as $\Hamiltonian_0$.
	Evaluate the ac driving matrix $V$ in the same basis.
	Extend the semiclassical Hamiltonian to include photons, as in Eq.~(\ref{Eq:QuantumModel}). 
	Evaluate this fully quantum Hamiltonian $\Hamiltonian_\text{QM}$ in the basis $\{\ket{i,n}\}$, where $n$ refers to the number of single-mode photons of energy $h\!f$.
	\item Identify the nearly degenerate manifolds $g_n$ of dimension $d=\text{dim}(\Hamiltonian_0)$ within the fully quantum Hamiltonian, as sketched in Fig.~\ref{Fig:EnergyLevel}.
	Block diagonalize $\Hamiltonian_\text{QM}$ by applying a Schrieffer-Wolff transformation to desired order~\cite{WinklerBook}, as in Eq.~(\ref{Eq:Hblock}).
	This yields a $d$-dimensional Hamiltonian $\tilde \Hamiltonian_n=\tilde \Hamiltonian_\text{block}+n h\!f$ corresponding to the perturbed manifold $\tilde{g}_n$, formed within the perturbed basis set $\{\ket{\tilde{i},n}\}$.
	Here $\tilde \Hamiltonian_\text{block}$ is independent of the photon number.
	\item Construct the $d$-dimensional time-evolution operator $\tilde U_n(t)$ for manifold $\tilde{g}_n$.
	Since $\tilde \Hamiltonian_\text{block}$ is independent of $n$, the time-evolution operators are also identical for each manifold, except for the phase factors $e^{-i n 2 \pi f\!t}$.
	\item Transform the time-evolution operator back to the original basis $\{\ket{i,n}\}$, yielding $\tilde U_\text{QM}(t)$.
	The correspondence between the quantum and semiclassical evolution operators is finally given by
\begin{equation}
U_\text{semi}(t)  =\bra{\alpha(t)}\tilde U_\text{QM}\ket{\alpha(0)}, 
\end{equation}
where $\alpha$ is the coherent state defined in Eq.~(\ref{Eq:CoherentState}).
$U_\text{semi}$ describes the full dynamics of the gate operation in the basis $\{\ket{0},\ket{1},\ket{L},\dots\}$.
\end{enumerate}

\subsection{Example: Two-level System With Transverse Drive}\label{Appendix:two-level}
In this section, we demonstrate the dressed-state formalism by applying it to a simple two-level system.
To take an example, we consider a charge qubit with constant tunnel coupling $\Delta$ and detuning parameter $\varepsilon$.
The ac drive $\varepsilon_\text{ac}=-2A\cos (2\pi f\!t)$ is applied to the detuning parameter.
(Here, the prefactor $-2$ is adopted for notational convenience.)
We assume that $\varepsilon$ has average value of $\bar\varepsilon=0$, similar to Ref.~\cite{Kim2015}, corresponding to the ``sweet spot" of the charge qubit.
In the left-right basis $\{|L \rangle,|R \rangle\}$, the double-dot Hamiltonian is given by
\begin{equation}
\Hamiltonian_{L,R} =
\begin{pmatrix}
-\varepsilon_\text{ac}/2 & \Delta \\
\Delta & \varepsilon_\text{ac}/2
\end{pmatrix} .
\end{equation}  
Note that there are no leakage states in this example.
We now discuss each step of the dressed-state formalism, following the labelling scheme given above.

\subsubsection{Quantum Hamiltonian}
We first diagonalize the undriven ($A=0$) Hamiltonian by transforming to the basis $\{|0\rangle = (|L\rangle -|R\rangle)/\sqrt{2}, |1\rangle = (|L\rangle +|R\rangle)/\sqrt{2}\}$.
The resulting semiclassical Hamiltonian is given by
\begin{equation}
\Hamiltonian_\text{semi} = \Hamiltonian_0 + V\! \cos (2 \pi f \!t),
\label{Eq:TLSSemi}
\end{equation}
where $\Hamiltonian_0=-\Delta\sigma_z$.
Here, the driving term $V= A\sigma_x$ is transverse, and we identify the qubit energy levels as $E_0=-\Delta$ and $E_1=+\Delta$.

Next, we extend the quantum dot Hamiltonian to include photons, writing $\Hamiltonian_\text{QM}=\Hamiltonian_\text{dot}+\Hamiltonian_\text{ph}+V_\text{int}$.
Here, the uncoupled dot Hamiltonian is given by $\Hamiltonian_\text{dot} = \sum_{i=0,1} E_i |i\rangle\langle i| \otimes I_\text{ph}$, the uncoupled photon Hamiltonian is given by $\Hamiltonian_\text{ph} = I_\text{dot} \otimes h\!f a^{\dagger} a$, and the interaction term is defined as $V_\text{int} = V_\text{dot} \otimes (a^{\dagger} + a)$.
We determine the relation between $V_\text{dot}$ and $V$ through the semiclassical correspondance principle of Eq.~(\ref{Eq:Vclassical}):
\begin{align*}
\langle \alpha| V_\text{int} | \alpha\rangle 
=& V_\text{dot}(\alpha_0^* e^{i 2 \pi f\!t} + \alpha_0 e^{- i 2 \pi f\!t})\\
=& 2|\alpha_0|V_\text{dot} \cos( 2 \pi f\!t + \phi),
\end{align*}
where we use the definitions $a|\alpha\rangle = e^{-i 2 \pi f\!t} \alpha_0 |\alpha\rangle$ and $\alpha_0= |\alpha_0| e^{-i \phi}$.
Note here that the phase $\phi$ determines the phase of the driving term in Eq.~(\ref{Eq:TLSSemi}), which in turn determines the rotation axis of the resonant gate operation in the $x$-$y$ plane.
In experimental settings, by convention, we define $\phi=0$ at the first application of a resonant gate, which corresponds to an $X$-rotation.
In subsequent applications of the resonant gate, the phase $\phi$ can be modified to provide other rotation axes in the $x$-$y$ plane.
Henceforth in this work, we will set $\phi=0$ for simplicity, so that $\alpha_0=|\alpha_0|$.
Finally then, using Eq.~(\ref{Eq:Vclassical}), we make the identification $V_\text{dot} = V/2 \sqrt{N}$.
Since $N = \bra{ \alpha }a^{\dagger} a\ket{\alpha}$, we then have
\begin{equation}
\bra{ \alpha } \Hamiltonian_\text{QM}\ket{ \alpha} = \Hamiltonian_\text{semi} + Nh\!f .
\label{Eq:Correspondence}
\end{equation}

Finally, we evaluate $\Hamiltonian_\text{QM}$ in the $\{\ket{i,n}\}$ basis.
To simplify the calculation, we note that coherent states involve a superposition of many photon number states, $\ket{n}$; however the predominant modes occur in the range $n\in [N - \Delta N,N + \Delta N]$, where $\Delta N/N \ll 1$. 
We can show that this range is indeed very narrow by noting from Eqs.~(\ref{Eq:CoherentState}) and (\ref{Eq:N}) that the probability of being in a state $\ket{n}$ is given by
\begin{equation}
P(n) = |\langle n | \alpha \rangle |^2 =e^{-N} \frac{N^n}{n!},  \label{eq:nalpha}
\end{equation}
corresponding to a Poisson distribution with a peak at $n=N$, and a width $\Delta N$ defined by
\begin{equation}
(\Delta N)^2 = \langle \alpha | (\hat{N} - N)^2 |\alpha \rangle = N.
\end{equation}
The limit $N \gg 1$ is appropriate for gate-driven fields, yielding $\Delta N / N = 1/\sqrt{N} \ll 1$.
We may therefore simplify the following calculations by replacing $n\rightarrow N$.

In this way, we obtain
\begin{equation}
\begin{aligned}
\langle i, m |V_\text{dot} a^{\dagger}|j,n\rangle &=  \langle j, m | V_\text{dot} \sqrt{n+1}|i,n+1 \rangle \\
&\simeq \langle i |V_\text{dot}| j \rangle |\alpha_0| \delta_{m,n+1} \\
&= \frac{V_{ij}}{2} \delta_{m,n+1},
\end{aligned}
\end{equation}
and similarly,
\begin{equation}
\langle i ,m |V_\text{dot} a |j,n\rangle \simeq \frac{V_{ij}}{2}  \delta_{m,n-1}.
\end{equation}
The individual terms in $\Hamiltonian_\text{QM}$ can then be expressed as 
\begin{gather}
\langle i, n |\Hamiltonian_\text{dot}|j, m \rangle = E_i\,\delta_{i,j}\delta_{n,m} , \label{eq:Hdotmatrix} \\
\langle i, n |\Hamiltonian_\text{ph}|j, m \rangle = n h\!f \, \delta_{i,j}\delta_{n,m} , \\
\langle i, n |V_\text{int}|j, m\rangle  = \frac{V_{ij}}{2} (\delta_{n,m+1}+ \delta_{n,m-1}) \label{eq:Vintmatrix} .
\end{gather}
Equations~(\ref{eq:Hdotmatrix})-(\ref{eq:Vintmatrix}) describe a band matrix with ``tri-block-diagonal'' form.
These general results apply to any driven two-level system, and do not depend specifically on the control parameter being driven.
For the case of a transversely driven Hamiltonian, as described above, we have $V_{01}=V_{10}= A$ and $V_{00}=V_{11}=0$.

To summarize this subsection, we have extended the semiclassical Hamiltonian of Eq.~(\ref{Eq:SemiHamiltonian}) to a full quantum model given by Eq.~(\ref{Eq:QuantumModel}) for the two-level system spanned by $i=0,1$.
Although $\Hamiltonian_\text{QM}$ is infinite-dimensional, it is instructive to write out a small portion of the full matrix.
For the basis states $S=\{|0,n-1\rangle,|1,n-1\rangle,|0,n\rangle,|1,n\rangle,|0,n+1\rangle,|1,n+1\rangle\}$, we have 
\begin{widetext}
\begin{equation}
\Hamiltonian_{S} = 
\begin{pmatrix}
E_0 + (n-1)h \!f & 0 & 0 & A/2 & 0 & 0 \\
0 & E_1 + (n-1)h \!f & A/2 & 0  & 0 & 0\\
0 & A/2 & E_0 + nh \!f & 0 & 0 & A/2  \\ 
A/2 & 0 & 0 & E_1 + nh \!f  & A/2 & 0 \\ 
0 & 0 & 0 & A/2 & E_0 + (n+1)h \!f & 0 \\ 
0 & 0 & A/2 & 0 & 0 & E_1 + (n+1)h \!f 
\end{pmatrix}. \label{eq:HS6D}
\end{equation}
\end{widetext}
Here we see that the Hamiltonian is sparse, since the $V_\text{int}$ only changes the photon number by one: $\langle i,n |\Hamiltonian_\text{QM}|j, m\rangle = 0$ when $ |n-m| \geq 2$.
As noted above, we are mainly interested in the portion of $\Hamiltonian_\text{QM}$ with $n\simeq N$.

\subsubsection{Block Diagonalization of the Dressed-State Hamiltonian}
In this step, we first identify the nearly degenerate $g_n$ manifolds of $\Hamiltonian_\text{QM}$, as illustrated in Fig.~\ref{Fig:EnergyLevel}, whose widths and separations are defined as $E_\text{intra}$ and $E_\text{inter}$.
We assume the system is driven near its single-photon resonance condition, defined as $h\!f\simeq E_1-E_0$.
The appropriate choice is $g_n=\{\ket{0,n+1},\ket{1,n}\}$, where
\begin{equation}
 |(E_1 + n h \!f)-(E_0 + (n+1) h \!f)| = E_\text{intra} < h\!f,
\label{Eq:NearlyDegenerateCondition}
\end{equation}
while
\begin{equation}
 |(E_1 + n h \!f)-(E_0 + (m+1) h \!f)| > E_\text{inter}, \,\, (n\neq m) .
\end{equation}
with $E_\text{inter}=h \!f -E_\text{intra} $.

The term $V_\text{int}$ provides the coupling between different $g_n$ manifolds, as indicated in Eq.~(\ref{eq:HS6D}).
We now block diagonalize $\Hamiltonian_\text{QM}$ into the perturbed manifolds $\tilde g_n=\{\ket{\tilde 0,n+1},\ket{\tilde 1,n}\}$ using the Schrieffer-Wolff decomposition method~\cite{WinklerBook, Orders}.
To second order in the small parameter $A/h\!f$, we obtain Eq.~(\ref{Eq:Hblock}), with
\begin{multline}
\tilde \Hamiltonian_n =\tilde \Hamiltonian_\text{block} + n h\! f  \\
=  \begin{pmatrix}
E_0 + (n+1)h\! f - \beta & A/2 \\
A/2& E_1 + n h\! f + \beta
\end{pmatrix} ,
\label{Eq:BlockHamiltonian}
\end{multline}
and
\begin{gather}
|\tilde{0}, n+1 \rangle = |0,n+1\rangle  - \gamma |1,n+2\rangle , \label{eq:0tilde} \\
\hspace{-.2in} |\tilde{1}, n \rangle = |1,n\rangle  + \gamma |0,n-1\rangle . \label{eq:1tilde}
\end{gather}
At this level of approximation, the the energy level shifts due to strong driving are given by $\pm\beta$, where $\beta = (A/2)^2 (E_1-E_0 + h\! f)^{-1}$, and the manifold hybridization factor is given by $\gamma = (A/2) (E_1-E_0 + h\! f)^{-1}$.
These represent leading order corrections to the RWA; additional corrections can be obtained, if desired,  by applying the Schrieffer-Wolff approximation to higher orders.
Resonance occurs when the diagonal elements of $\tilde\Hamiltonian_n$ are equal:  $(\tilde\Hamiltonian_n)_{00} = (\tilde\Hamiltonian_n)_{11}$.
Hence,
\begin{equation}
\label{Eq:BlochSiegertTLS}
h\!\tilde f_\text{res} = E_1-E_0 + 2 \beta ,
\end{equation}
where $h\! f_0 = E_1-E_0$ is the bare resonant frequency, and we identify $2\beta$ as the Bloch-Siegert shift.
As is well known \cite{Bloch1940,DressedAtom}, the energy denominator in $\beta$ is given approximately by $2h\!f$.
Here, the factor of 2 occurs because the term arises from the counter-rotating term in the drive.

\subsubsection{Time Evolution of Manifold $\tilde{g}_n$}
Since $\tilde \Hamiltonian_\text{block}$ is time-independent and the manifolds are decoupled, the evolution operator for manifold $\tilde g_n$ is simply given by
\begin{equation}
\begin{aligned}
\tilde U_n(t) =& \,\, \exp\left[-\frac{i}{\hbar} (\tilde \Hamiltonian_\text{block} + n h\! f)t\right] \\
=& \,\, \exp\left[-\frac{i}{\hbar}\left(\frac{E_0 + h\! f +E_1}{2} + n h\! f\right)\!t\right] \\
&\hspace{-.45in}\times[\cos(\pi f_\text{Rabi}t) - i \sin (\pi f_\text{Rabi}t) 
(\sigma_{zn}\cos \theta  + \sigma_{xn} \sin \theta )], 
\end{aligned}
\label{Eq:Un}
\end{equation}
where 
\begin{gather}
f_\text{Rabi} = \frac{2}{h} \sqrt{\left[\frac{(E_0+h\! f)-E_1}{2}-\beta\right]^2 + \left[\frac{A}{2}\right]^2} , \label{Eq:fRabiTLS}\\
\cos \theta =\frac{1}{h\! f_\text{Rabi}} \left[(E_0 + h\! f) -E_1-2 \beta\right] , \label{Eq:CosTLS}\\
\sin \theta =\frac{A}{h\! f_\text{Rabi}} , \label{Eq:SinTLS}
\end{gather}
and the Pauli operators $\sigma_{zn}$ and $\sigma_{xn}$ refer specifically to the $\tilde g_n$ manifold, whose basis states are given by Eqs.~(\ref{eq:0tilde}) and (\ref{eq:1tilde}).
Note that when the qubit is driven resonantly at $f=\tilde f_\text{res}$, the Rabi frequency reduces to $h\!f_\text{Rabi}=A$, representing the standard Rabi result.  
Hence, the Rabi frequency does not acquire any strong-driving corrections at this level of approximation.

\subsubsection{Semi-classical Evolution Operator}
We now evaluate the evolution operator in the bare-state basis, $\{\ket{i,n}\}$.
We proceed by inverting Eqs.~(\ref{eq:0tilde}) and (\ref{eq:1tilde}) and assuming resonant driving, yielding
\begin{widetext}
\begin{eqnarray}
\tilde U_\text{QM}|0,n+1\rangle 
&=&  \tilde U_\text{QM} (|\tilde{0},n+1\rangle + \gamma |\tilde{1}, n+2\rangle) 
=  \tilde U_n|\tilde{0},n+1\rangle + \gamma \tilde U_{n+2}| \tilde{1}, n+2\rangle \nonumber \\
&=&  e^{-\frac{i}{\hbar}\left(\frac{E_0 + h\! \tilde f_\text{res} +E_1}{2}+n h\! \tilde f_\text{res} \right)t}
[\cos (\Omega t/2) |\tilde{0}, n+1\rangle - i \sin (\Omega t/2) |\tilde{1},n\rangle]  \nonumber \\
&& \hspace{0.15 in} +  \gamma e^{-\frac{i}{\hbar}\left(\frac{E_0 + h\! \tilde f_\text{res} +E_1}{2}+ (n+2) h\! \tilde f_\text{res} \right)t} 
[\cos(\Omega t/2) |\tilde{1}, n+2\rangle - i \sin (\Omega t/2) |\tilde{0}, n+3 \rangle]  \nonumber \\
&=&  e^{-\frac{i}{\hbar}\left(\frac{E_0 + h\! \tilde f_\text{res}  +E_1}{2}+n h\! \tilde f_\text{res} \right)t} 
\left[\cos(\Omega t/2) |0, n+1\rangle 
- i \sin (\Omega t/2) |1,n\rangle 
   \right.  - \gamma (1-e^{- i 4 \pi \tilde f_\text{res} t})\cos(\Omega t/2) |1, n+2\rangle \nonumber \\
&& \hspace{0.15 in} \left.
- i \gamma \sin (\Omega t/2) |0, n-1 \rangle 
-  i\gamma  e^{-i 4 \pi \tilde f_\text{res} t} \sin (\Omega t/2) |0, n+3 \rangle 
\right] ,
\label{Eq: Un1}
\end{eqnarray}
where $\Omega = 2 \pi f_\text{Rabi}$.  Similarly,

\begin{eqnarray}
\tilde U_\text{QM}|1,n\rangle 
&=&  e^{-\frac{i}{\hbar}\left(\frac{E_0 + h\! \tilde f_\text{res} +E_1}{2}+n h\! \tilde f_\text{res} \right)t} 
\left[ \cos(\Omega t/2) |1, n\rangle - i \sin (\Omega t/2) |0,n+1\rangle  \right. \label{Eq: Un2}  \\
&& \hspace{0.15 in} \left. 
+ \gamma (1-e^{ i 4 \pi \tilde f_\text{res} t})\cos(\Omega t/2) |0, n-1\rangle 
+  i \gamma \sin (\Omega t/2) |1, n+2 \rangle + i\gamma  e^{i 4 \pi \tilde f_\text{res} t} \sin (\Omega t/2) |1, n-2 \rangle  
\right] . \nonumber
\end{eqnarray}
\end{widetext}
In Eqs.~(\ref{Eq: Un1}) and (\ref{Eq: Un2}), we note that if the system is initially in a state with a fixed photon number, then over time it will diffuse into many different photon states.  
However, we now show that if the initial state is a coherent state, then it will remain in the same time-evolved state.  (Indeed, coherent states are designed to have this property for large $N$, due to their correspondence to classical fields \cite{DressedAtom}.)

The coherent state at $t=0$ can be written as 
\begin{equation}
|\alpha(t=0)\rangle = \sum\limits_n c_n |n\rangle ,
\end{equation}
where from Eq.~(\ref{Eq:CoherentState}), we have 
\begin{equation}
c_n=e^{-N/2}\sqrt{N^n/n!} \,\, . \label{eq:cn}
\end{equation}
Here again, we have chosen the phase of $\alpha_0$ such that $\phi=0$.
In the bare-state basis, the time-evolution operator can be expressed in the general form
\begin{equation}
\tilde U_\text{QM} (t)|i,n\rangle = e^{-i n 2 \pi f\! t} \sum\limits_{j,m} B_{ij,m} |j, n+m \rangle .
\end{equation}
For the  current example, the tensor $B_{ij,m}$, corresponding to Eqs.~(\ref{Eq: Un1}) and (\ref{Eq: Un2}), is
\begin{align*}
B_{00,0} &= e^{-\frac{i}{\hbar}\left[\frac{E_0 - h\! \tilde f_\text{res} +E_1}{2}\right]t} \cos(\Omega t/2) ,\\
B_{00,-2} &= - i e^{-\frac{i}{\hbar}\left[\frac{E_0 - h\! \tilde f_\text{res} +E_1}{2}\right]t} \gamma \sin (\Omega t/2) ,\\
B_{00,2} &= -i e^{-\frac{i}{\hbar}\left[\frac{E_0 - h\! \tilde f_\text{res} +E_1}{2}\right]t}\gamma e^{-i 4 \pi \tilde f_\text{res} t} \sin (\Omega t/2) ,\\
B_{01,-1} &=  - i e^{-\frac{i}{\hbar}\left[\frac{E_0 - h\! \tilde f_\text{res} +E_1}{2}\right]t}  \sin (\Omega t/2) ,\\
B_{01,1} &=  - e^{-\frac{i}{\hbar}\left[\frac{E_0 - h\! \tilde f_\text{res} +E_1}{2}\right]t}  \gamma (1-e^{- i 4 \pi \tilde f_\text{res} t})\cos(\Omega t/2) ,\\
B_{10,-1} &=   e^{-\frac{i}{\hbar}\left[\frac{E_0 + h\! \tilde f_\text{res} +E_1}{2}\right]t}  \gamma (1-e^{i 4 \pi \tilde f_\text{res} t})\cos(\Omega t/2) , \\
B_{10,1} &=  - i e^{-\frac{i}{\hbar}\left[\frac{E_0 + h\! \tilde f_\text{res} +E_1}{2}\right]t}  \sin (\Omega t/2) , \\
B_{11,0} &=   e^{-\frac{i}{\hbar}\left[\frac{E_0 + h\! \tilde f_\text{res} +E_1}{2}\right]t}  \cos(\Omega t/2) , \\
B_{11,-2} &=   i e^{-\frac{i}{\hbar}\left[\frac{E_0 + h\! \tilde f_\text{res} +E_1}{2}\right]t}\gamma e^{i 4 \pi \tilde f_\text{res} t} \sin (\Omega t/2) , \\
B_{11,2} &=   i e^{-\frac{i}{\hbar}\left[\frac{E_0 + h\! \tilde f_\text{res} +E_1}{2}\right]t} \gamma \sin (\Omega t/2) . \\
\end{align*}

Using Eqs.~(\ref{Eq:CoherentState}) and (\ref{eq:Usemi}), it is now easy to show that the semiclassical time evolution is given by
\begin{align}
U_\text{semi}\ket{i} =& \bra{\alpha(t)}\tilde{U}_\text{QM}\ket{i,\alpha(0)} \nonumber \\
=&\sum_{n,m,j}|c_n|^2\left|\frac{c_{n+m}}{c_n}\right|e^{i 2\pi m\tilde f_\text{res} t}B_{ij,m}\ket{j} .
\end{align}
Using Stirling's approximation and Eq.~(\ref{eq:cn}), it is also easy to show that

\begin{equation}
\left|\frac{c_{n+m}}{c_n}\right| \simeq \exp \left[-\frac{m^2 + 2 m (n-N)}{4N}\right] \simeq 1 ,
\end{equation}
where we have taken $n\simeq N$ and $m\ll N$.  
Finally, noting that $\sum |c_n|^2 =1$, we obtain
\begin{equation}
U_\text{semi}\ket{i} =\sum_{m,j}e^{i 2\pi m \tilde f_\text{res} t}B_{ij,m}\ket{j} .
\end{equation}

\begin{figure}
\includegraphics[width=3.2in]{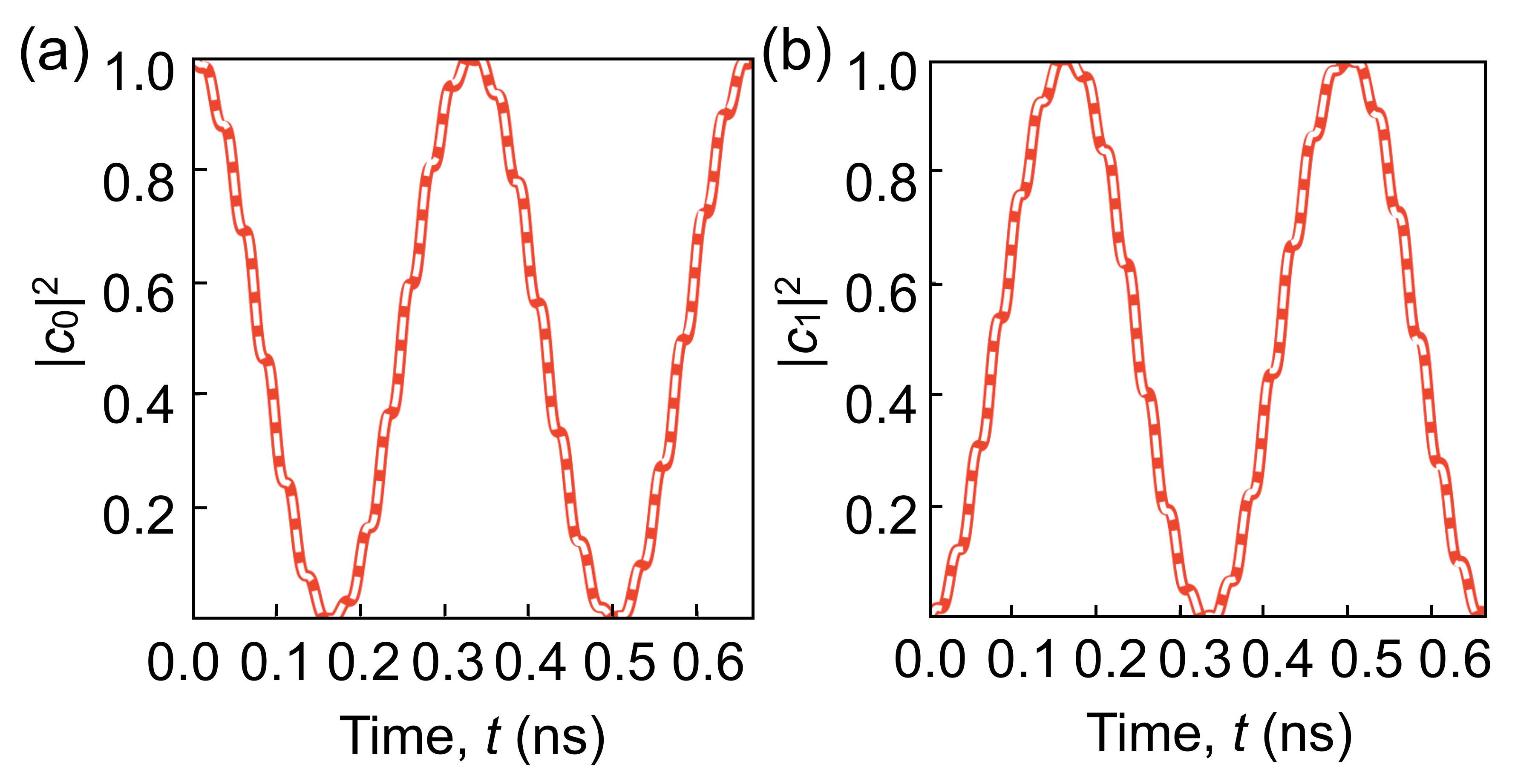}
\caption{
The dynamical evolution of a transversely driven two-level system under strong resonant driving.
Exact numerical results are shown as thick red lines, while the analytical results obtained in Eqs.~(\ref{Eq:TLSresult1}) and (\ref{Eq:TLSresult2}) are shown as thin white lines; the square of the difference between the two solutions is less than $10^{-4}$ over the whole range of the plot.
Here, we take $\ket{0}$ as the initial state and plot the probabilities $P_0=|c_0|^2$ and $P_1=|c_1|^2$ of being in qubit states $\ket{0}$ and $\ket{1}$ as a function of time.
The system parameters are given by $\{E_0, E_1, A\}/h = \{-10,10,3\}$~GHz, where $E_0$ and $E_1$ are the energies of the qubit states and $A$ is the amplitude of the drive.
}
\label{Fig:TLS}
\end{figure}

Finally, we obtain the time evolution of the transversely driven two-level system, driven on resonance:
\begin{eqnarray}
&&U_\text{semi}\ket{0} =  \label{Eq:TLSresult1}  \\
&& \hspace{.3in} \cos (\Omega t/2)\ket{0}-i\sin (\Omega t/2)e^{-i2\pi \tilde f_\text{res} t}\ket{1}  \nonumber \\
&& \hspace{.3in}
-2i\gamma [\cos(\Omega t/2)\sin (2\pi \tilde f_\text{res} t)\ket{1} \nonumber  \\
&& \hspace{.3in}
+\sin(\Omega t/2)\cos(2\pi \tilde f_\text{res} t)e^{-i2\pi \tilde f_\text{res} t}\ket{0} ] , \nonumber \\
&&U_\text{semi}\ket{1} =  \label{Eq:TLSresult2} \\
&& \hspace{.3in}
e^{-i 2 \pi \tilde f_\text{res} t} \left( \cos (\Omega t/2)\ket{1}-i\sin (\Omega t/2)e^{i2\pi \tilde f_\text{res} t}\ket{0}  \right.\nonumber  \\
&& \hspace{.3in}
-2i\gamma [\cos(\Omega t/2)\sin (2\pi \tilde f_\text{res} t)\ket{0}\nonumber  \\
&& \hspace{.3in} 
-\left. \sin (\Omega t/2)\cos(2\pi \tilde f_\text{res} t)e^{i2\pi \tilde f_\text{res} t}\ket{1} ]\right) , \nonumber 
\end{eqnarray}
where we have dropped an overall phase term.
For each of these equations, the first line corresponds to the standard Rabi solution, while the second and third lines represents strong-driving corrections.
If we assume that the system is initially prepared in its ground state $\ket{0}$, this yields the following leading-order results for the probability evolution:
\begin{gather}
P_0(t)=|c_0(t)|^2=\cos^2(\Omega t/2)-\gamma\sin(\Omega t)\sin(4\pi \tilde f_\text{res} t) \label{eq:P0} ,\\
P_1(t)=|c_1(t)|^2=\sin^2(\Omega t/2)+\gamma\sin(\Omega t)\sin(4\pi \tilde f_\text{res} t) \label{eq:P1}.
\end{gather}
Here, we observe the emergence of fast oscillations with frequency $2\tilde f_\text{res}$ and amplitude given by $\gamma \sin(\Omega t)\propto A$.
Together with the renormalization of $f_\text{res}$ by the Bloch-Siegert shift, these represent the main effects of the counter-rotating term on the evolution of the two-level system.

Figure~\ref{Fig:TLS} shows a comparison of our analytical results, obtained above, and the corresponding numerical simulations of the evolution of a two-level system with a transverse drive.
We see that the analytical results provide an excellent description of the dynamics, including strong-driving effects.

To summarize:  the main (slow) oscillations in Fig.~\ref{Fig:TLS} represent the conventional Rabi results. The fine structure is due to the counter-rotating terms, which are dropped in the RWA, but which can have a strong effect on the gate fidelity when the drive is strong.  In principle, the dressed-state method captures all strong driving effects if we keep all the terms in $\Hamiltonian_\text{QM}$; however we can obtain corrections at increasing orders of approximation by block-diagonalizing larger subsets of the full dressed-state Hamiltonian,
or by including higher-order terms that arise in the block-diagonalization procedure. 
Finally, we note that the results obtained here were simplified under the assumption of resonant driving; however, more general, non-resonant results can also be obtained in the same manner.

\section{Dressed State Analysis of the Quantum Dot Hybrid Qubit }
\label{Sec:HybridQubitDetail}
We now provide the details of our main results for quantum dot hybrid qubits, which were summarized in Sec.~\ref{Sec:HybridQubitResult} of the main text.  
We first derive expressions for the driving matrix $V$ in the large-detuning regime.
We then derive the time-evolution operator $U_\text{semi}$ at lowest order (RWA) and next-lowest order in the expansion parameter $A/h\!f$, using the formalism described in Sec.~\ref{Sec:DressedStateFormalism} and Appendix~\ref{sec:procedure}.
As before, $A$ is the driving amplitude and $h\!f$ is the energy spacing between the triplet manifolds. 
Since analytical results are difficult to obtain, except in special cases, we focus below on the large-detuning limit.
However, we note that the simulations reported in this paper do not involve such approximations and are exact, up to numerical accuracy.

\subsection{Driving Matrix in the Large-detuning Regime}\label{sec:LargeDetuning}

\subsubsection{Tunnel Coupling Driving}\label{sec:HybridAppendix}
As consistent with Eq.~(\ref{Eq:Hamiltonian}), in the basis $\{\ket{\cdot S},\ket{\cdot T},\ket{S\cdot}\}$, the time-dependent, semi-classical Hamiltonian with tunnel coupling driving is given by
\begin{equation}
\Hamiltonian = 
\begin{pmatrix}
-\frac{\varepsilon }{2} & 0 & \Delta_{1}(t) \\
0 & -\frac{\varepsilon}{2} + E_\text{ST} & -\Delta_{2}(t) \\
\Delta_{1}(t)  & -\Delta_{2}(t) & \frac{\varepsilon}{2}
\end{pmatrix} ,
\end{equation}
where $\Delta_i(t) = \Delta_{i0} + r_i \Delta_\text{ac}(t)$ for $i=1,2$ and $\Delta_\text{ac}(t) = A \cos (2 \pi f t)$.
We now consider the far-detuned limit $\varepsilon\gg \Delta_1,\Delta_2,E_\text{ST}$ and diagononalize the undriven Hamiltonian $\Hamiltonian_0$ up to leading order in the small parameter $\Delta_i/\varepsilon$, yielding the eigenbasis $\{|0\rangle,|1\rangle,|L\rangle\}$ and the corresponding energies
\begin{gather}
E_0\simeq -\frac{\varepsilon}{2}-\frac{\Delta_1^2}{\varepsilon} , \label{eq:E0AppC} \\
E_1\simeq -\frac{\varepsilon}{2}+E_\text{ST}-\frac{\Delta_2^2}{\varepsilon-E_\text{ST}} , \label{eq:E1AppC} \\
E_L\simeq \frac{\varepsilon}{2}+\frac{\Delta_1^2}{\varepsilon} +\frac{\Delta_2^2}{\varepsilon-E_\text{ST}}  , \label{eq:ELAppC}
\end{gather}
which are consistent with Eq.~(\ref{Eq:Heff}) in the appropriate limit.
We then evaluate the driving term in this basis, obtaining $\Hamiltonian = \Hamiltonian_0 + V \cos (2 \pi f t)$, where
\begin{equation}
V\simeq A\!\!
\begin{pmatrix}
-\frac{2 \Delta_{10} r_{1}}{\varepsilon} & \frac{\Delta_{10} r_{2}}{\varepsilon} \!+\!\frac{\Delta_{20} r_{1}}{\varepsilon-E_\text{ST}} & r_1  \\
\frac{\Delta_{10} r_{2}}{\varepsilon} \!+\!\frac{\Delta_{20} r_{1}}{\varepsilon-E_\text{ST}} & -\frac{2 \Delta_{20} r_{2}}{\varepsilon -E_\text{ST}} &  -r_2 \\
r_1  & -r_2 & \frac{2 \Delta_{10} r_{1}}{\varepsilon}\!+\!\frac{2 \Delta_{20} r_{2}}{\varepsilon -E_\text{ST}}
\end{pmatrix} .
\label{Eq:Vtunnel}
\end{equation}
In particular, we see that $V_{01}\simeq A[\Delta_{10} r_{2}/\varepsilon + \Delta_{20} r_{1}/(\varepsilon-E_\text{ST})]$, which gives the leading order (RWA) expression for the Rabi frequency, $h\!f_\text{Rabi}=V_{01}$, when the qubit is driven on resonance, as consistent with Eq.~(\ref{eq:fRabifres}).

\subsubsection{Detuning Driving}
Similarly, in the basis $\{\ket{\cdot S},\ket{\cdot T},\ket{S\cdot}\}$, the semi-classical Hamiltonian for detuning driving is given by
\begin{equation}
\Hamiltonian = 
\begin{pmatrix}
-\frac{\varepsilon + \varepsilon_\text{ac}(t)}{2} & 0 & \Delta_{10}  \\
0 & -\frac{\varepsilon + \varepsilon_\text{ac}(t)}{2} + E_\text{ST} & -\Delta_{20} \\
\Delta_{10}  & -\Delta_{20} & \frac{\varepsilon + \varepsilon_\text{ac}(t)}{2}
\end{pmatrix} ,
\end{equation}
where $\varepsilon_\text{ac}(t) = A \cos (2 \pi f t)$.
Again assuming the far-detuned limit, we obtain the Hamiltonian $\Hamiltonian = \Hamiltonian_0 + V \cos (2 \pi f t)$ in the $\{|0\rangle,|1\rangle,|L\rangle\}$ basis, with energies given by Eqs.~(\ref{eq:E0AppC})-(\ref{eq:ELAppC}), and
\begin{eqnarray}
V &\simeq& \label{Eq:Vdetuning} \\
&& \hspace{-.2in} \nonumber
A\begin{pmatrix}
-\frac{1}{2} +\frac{\Delta_1^2}{\varepsilon^2} & -\frac{\Delta_{1} \Delta_{2}}{\varepsilon (\varepsilon-E_\text{ST})} & -\frac{\Delta_{1}}{\varepsilon}  \\
-\frac{\Delta_{1} \Delta_{2}}{\varepsilon (\varepsilon-E_\text{ST})} & -\frac{1}{2} +\frac{\Delta_2^2}{(\varepsilon-E_\text{ST})^2}&  \frac{\Delta_{2}}{\varepsilon-E_\text{ST}} \\
-\frac{\Delta_{1}}{\varepsilon}  & \frac{\Delta_{2}}{\varepsilon-E_\text{ST}} & \frac{1}{2} -\frac{\Delta_1^2}{\varepsilon^2}-\frac{\Delta_2^2}{(\varepsilon-E_\text{ST})^2}
\end{pmatrix} .
\end{eqnarray}
In this case, we see that $V_{01}\simeq -A\Delta_1\Delta_2/\varepsilon(\varepsilon-E_\text{ST})$.

\subsection{Lowest-Order Results}
We now derive the lowest-order (RWA) results for the quantum dot hybrid qubit, assuming that $\varepsilon > 2.5 E_\text{ST}$~\cite{Orders}.
In this regime, we have $E_1-E_0\simeq E_\text{ST}$ and $E_L-E_0\simeq \varepsilon$. 
Note that the driving matrix $V$ is not specified here -- it can describe tunnel coupling driving [Eq.~(\ref{Eq:Vtunnel})], detuning driving [Eq.~(\ref{Eq:Vdetuning})], or even a combination of the two.
 
Following our prescription in Appendix~\ref{sec:procedure} for constructing dressed states in a three-level system, we obtain results that are lowest-order in $A/hf$ for the perturbed Hamiltonian $\tilde\Hamiltonian_n$, which acts on the manifold $\tilde g_n=\{\ket{\tilde 0, n+1},\ket{\tilde 1, n},\ket{\tilde L, n-k}\}$.  The corresponding block Hamiltonian is given by
\begin{equation}
\tilde \Hamiltonian_\text{block} = 
\begin{pmatrix}
E_0 + h\! f & V_{01}/2 & 0 \\
V_{01}/2 & E_1 & 0 \\
0 & 0 & E_L - k h\! f
\end{pmatrix} . \label{eq:HnRWA}
\end{equation}
At this level of approximation, we obtain $h\! f_0 = E_1-E_0$ for the RWA resonance frequency and $h\! f_\text{Rabi} = V_{01} = \hbar \Omega$ for the Rabi frequency. 

In principle, the evolution operator for an arbitrary driving frequency $f$ can be derived.
Here, for simplicity, we assume resonant driving, $f = f_0$; then the evolution operator in the basis $\{\ket{0},\ket{1},\ket{L}\}$ is given by
\begin{eqnarray}
U_\text{semi} &=&  \label{eq:UsemiRWA} \\
&& \hspace{-0.45in} \nonumber
\begin{pmatrix}
e^{-\frac{i}{\hbar} E_0 t} \cos (\Omega t/2) & - i e^{-\frac{i}{\hbar} E_0 t} \sin (\Omega t/2) & 0\\
- i e^{-\frac{i}{\hbar} E_1 t} \sin (\Omega t/2) & e^{-\frac{i}{\hbar} E_1 t} \cos (\Omega t/2) & 0 \\
0 & 0 & e^{- \frac{i}{\hbar} E_L t} 
\end{pmatrix}. 
\end{eqnarray}
We note that this operator includes no coupling to the leakage state.
There are two reasons for this decoupling.
First, since $\varepsilon > 2.5 E_\text{ST}$, we must have $k \geq 2$; however $V_\text{int}$ only couples states that differ by one photon.
Second, the hybridization of dot states is unimportant at this level of approximation.
As a result, the leakage state in Eq.~(\ref{eq:UsemiRWA}) remains decoupled for all times.
(This is not true for $\varepsilon<2.5E_\text{ST}$, however.)
Within the logical subspace $\{\ket{0},\ket{1}\}$, Eq.~(\ref{eq:UsemiRWA}) thus describes the conventional Rabi result, and does not include any strong-driving corrections.

\subsection{Second-Order Results}
We now present next-order results for the quantum dot hybrid qubit, assuming $\varepsilon > 3.5 E_\text{ST}$. 
At this order, we obtain
\begin{equation}
\tilde \Hamiltonian_\text{block} = 
\begin{pmatrix}
E_0 + h\! f + \beta_0 & V_{01}/2 & 0 \\
V_{01}/2 & E_1 + \beta_1 & 0 \\
0 & 0 & E_L - k h\! f  + \beta_L
\end{pmatrix} ,
\label{Eq:HblockSecond}
\end{equation}
where
\begin{widetext}
\begin{gather}
\beta_0 =  \frac{V_{01}^2}{4 (E_0-E_1 -h\! f)}  + \frac{V_{0L}^2}{4 (E_0-E_L - h\! f)} + \frac{V_{L0}^2}{4 (E_0-E_L + h\! f)} ,\\
\beta_1 =  \frac{V_{01}^2}{4 (-E_0+E_1 +h\! f)}  
+ \frac{V_{1L}^2}{4 (E_1-E_L -h\! f)} + \frac{V_{L1}^2}{4 (E_1-E_L +h\! f)} ,\\
\beta_L =  \frac{V_{0L}^2}{4 (-E_0+E_L +h\! f)} + \frac{V_{L0}^2}{4 (-E_0+E_L -h\! f)} 
+ \frac{V_{1L}^2}{4 (-E_1+E_L +h\! f)} + \frac{V_{L1}^2}{4 (-E_1+E_L -h\! f)} 
\end{gather}
\end{widetext}
are the Bloch-Siegert shifts.
The resonant driving frequency is now given by $h\! \tilde f_\text{res} = E_1 + \beta_1 - E_0 -\beta_0$, while the Rabi oscillation frequency is still given by $h\! f_\text{Rabi} = V_{01} = \hbar \Omega$.

At this order, the perturbed manifold is given by
\begin{widetext}
\begin{eqnarray}
|\tilde 0, n+1\rangle &=& |0, n+1\rangle +\frac{V_{00}}{2h\!f}|0,n\rangle - \frac{V_{00}}{2h\!f}|0,n+2\rangle - \frac{V_{01}}{2(E_1-E_0 + h\!f)}|1,n+2\rangle \label{Eq:hybridstate0} \\
&& \hspace{-0.3 in}
 - \frac{V_{0L}}{2 (E_L-E_0 - h\!f)}|L,n\rangle - \frac{V_{0L}}{2(E_L-E_0 + h\!f)}|L,n+2\rangle \nonumber \\
|\tilde 1, n\rangle &=& |1, n\rangle +\frac{V_{10}}{2 (E_1-E_0+h\!f)}|0,n-1\rangle + \frac{V_{11}}{2h\!f}|1,n-1\rangle - \frac{V_{11}}{2h\!f}|1,n+1\rangle \label{Eq:hybridstate1} \\
&& \hspace{-0.3 in} 
 - \frac{V_{1L}}{2 (E_L-E_1-h\!f)}|L,n-1\rangle - \frac{V_{1L}}{2(E_L-E_1 + h\!f)}|L,n+1\rangle \nonumber \\
|\tilde L, n-k\rangle &=& |L, n-k\rangle + \frac{V_{L0}}{2(E_L - E_0 + h\!f)} |0, n-k-1\rangle + \frac{V_{L0}}{2(E_L - E_0 - h\!f)} |0, n-k+1\rangle   \label{Eq:hybridstateL} \\
&& \hspace{-0.3 in}
+\frac{V_{L1}}{2(E_L - E_1 + h\!f)}|1,n-k-1\rangle + \frac{V_{L1}}{2(E_L - E_1 - h\!f)}|1,n-k+1\rangle + \frac{V_{LL}}{2h\!f} |L, n-k-1\rangle - \frac{V_{LL}}{2h\!f}|L, n-k+1\rangle  \nonumber 
\end{eqnarray}

Again, for simplicity, we assume resonant driving at $f = \tilde f_\text{res}$. 
The matrix elements of $U_\text{semi}$ are then given by
\begin{eqnarray}
\left(U_\text{semi}\right)_{00} &= & \, e^{-\frac{i}{\hbar} (E_0 + \beta_0 ) t} \cos (\Omega t/2) - 
i \frac{ V_{00} \sin (2 \pi \tilde f_\text{res} t) \, e^{-\frac{i}{\hbar} (E_0 + \beta_0 ) t}}{h f_0}  \cos (\Omega t/2) 
\nonumber \\ && \hspace{0.3in}
-i \frac{V_{01} \cos (2 \pi \tilde f_\text{res} t) \,  e^{-\frac{i}{\hbar} (E_1 +  \beta_1 ) t}}{\red{E_1-E_0 + h f_0}} \sin (\Omega t/2)  ,
\\ \left(U_\text{semi}\right)_{10} &= &  - i e^{-\frac{i}{\hbar} (E_1 + \beta_1) t} \sin (\Omega t/2)  -  
\frac{V_{11} \sin (2 \pi \tilde f_\text{res} t) \,  e^{-\frac{i}{\hbar} (E_1 + \beta_1) t}}{h f_0}  \sin (\Omega t/2)  
\nonumber \\ && \hspace{0.3in}
- i \frac{ V_{01} \sin (2 \pi \tilde f_\text{res} t) \, e^{-\frac{i}{\hbar} (E_0 + \beta_0) t}}{\red{E_1-E_0 + h f_0}}  \cos (\Omega t/2)   ,
\\ \left(U_\text{semi}\right)_{01} & = &  \, - i e^{-\frac{i}{\hbar} (E_0 +  \beta_0) t} \sin(\Omega t/2) - 
 \frac{V_{00}\sin (2 \pi \tilde f_\text{res} t) \,  e^{-\frac{i}{\hbar} (E_0 +  \beta_0 ) t}}{h f_0} \sin(\Omega t/2) 
 \nonumber \\ && \hspace{0.3in}
 - i \frac{ V_{01} \sin (2 \pi \tilde f_\text{res} t) \, e^{-\frac{i}{\hbar} (E_1 + \beta_1) t} }{\red{E_1-E_0 + h f_0}} \cos (\Omega t/2) ,
\\ \left(U_\text{semi}\right)_{11} & = & e^{-\frac{i}{\hbar} (E_1 + \beta_1) t} \cos (\Omega t/2) 
- i \frac{V_{11} \sin  (2 \pi \tilde f_\text{res} t) \, e^{-\frac{i}{\hbar} (E_1 + \beta_1) t}}{h f_0}  \cos(\Omega t/2)  
 \nonumber \\ && \hspace{0.3in}
+ i \frac{V_{01} \cos (2 \pi \tilde f_\text{res} t) \, e^{-\frac{i}{\hbar} (E_0 + \beta_0) t}}{\red{E_1-E_0 + h f_0}}   \sin (\Omega t/2)    ,
\end{eqnarray}
corresponding to the evolution within the logical subspace, $\{|0\rangle, |1\rangle\}$.  
The leakage state evolves according to
\begin{eqnarray}
\left(U_\text{semi}\right)_{L0} &=& \, \left( \frac{V_{0L} e^{ i 2 \pi \tilde f_\text{res} t} }{2 (E_0-E_L-h f_0)} + \frac{V_{L0} e^{ -i 2 \pi \tilde f_\text{res} t} }{2(E_0-E_L+ h f_0)}\right) e^{-\frac{i}{\hbar} (E_0 + \beta_0 ) t} \cos (\Omega t/2) 
\nonumber \\ && \hspace{0.3in}
-  i \left( \frac{ V_{1L} e^{i 2 \pi \tilde f_\text{res} t} }{2(E_1-E_L- h f_0)} + \frac{ V_{L1} e^{-i 2 \pi \tilde f_\text{res} t} }{2(E_1-E_L+ h f_0)}\right)  e^{-\frac{i}{\hbar} (E_1 + \beta_1) t} \sin (\Omega t/2)  
\nonumber \\ && \hspace{0.3in}
+ \left( \frac{V_{0L}}{2 (-E_0+E_L+ h f_0)} + \frac{V_{L0}}{2 (-E_0+E_L- h f_0)}\right) e^{- \frac{i}{\hbar} (E_L + \beta_L ) t} , \label{eq:UL0} \\
\left(U_\text{semi}\right)_{L1} &=& 
-i \left( \frac{V_{0L} e^{ i 2 \pi \tilde f_\text{res} t}}{2(E_0-E_L- h f_0)} + \frac{V_{L0} e^{ -i 2 \pi \tilde f_\text{res} t}}{2(E_0-E_L+h f_0)}\right) e^{-\frac{i}{\hbar} (E_0 +  \beta_0 ) t} \sin(\Omega t/2)  
\nonumber \\ && \hspace{0.3in}
+ \left( \frac{V_{1L}  e^{i 2 \pi \tilde f_\text{res} t} }{2(E_1-E_L-h f_0)} + \frac{ V_{L1}  e^{-i 2 \pi \tilde f_\text{res} t}}{2(E_1-E_L+h f_0)}\right) e^{-\frac{i}{\hbar} (E_1 + \beta_1) t} \cos (\Omega t/2)  
\nonumber \\ && \hspace{0.3in}
+ \left( \frac{V_{1L}}{2 (-E_1+E_L+h f_0)} + \frac{ V_{L1}}{2 (-E_1+E_L-h f_0)}\right) e^{- \frac{i}{\hbar} (E_L + \beta_L) t} . \label{eq:UL1}
\end{eqnarray}
\end{widetext}
For brevity here, we have omitted the terms $\left(U_\text{semi}\right)_{0L}$, $\left(U_\text{semi}\right)_{1L}$, and $\left(U_\text{semi}\right)_{LL}$, since they play a relatively minor role when the qubit is initialized into the logical subspace.

We can visualize these results more clearly by considering a qubit initialized into state $\ket{0}$.
In this case, to lowest order in $A$, the solution for the probabilities of being in the qubit states, $P_0$ and $P_1$, \red{are given by} 
\begin{eqnarray}
P_0 &=& |c_0|^2=\cos^2(\Omega t/2) \label{eq:P0hybrid} \\ \nonumber
&& \hspace{.1in} 
-\frac{V_{01}}{2(E_1-E_0+h\! f_0)}\sin(\Omega t)\sin(4\pi \tilde f_\text{res}t) , \\
P_1 &=& |c_1|^2=\sin^2(\Omega t/2) \label{eq:P1hybrid} \\ \nonumber
&& \hspace{.1in} 
+\frac{V_{01}}{2(E_1-E_0+h\! f_0)}\sin(\Omega t)\sin(4\pi \tilde f_\text{res}t) .
\end{eqnarray}
\red{The solution for the leakage state, $P_L$, is oscillatory and proportional to $A^2$; however, its form is rather complicated and we omit it here for brevity. 
}
Similar to Eqs.~(\ref{eq:P0}) and (\ref{eq:P1}) for the two-level system, we again observe Rabi oscillations at frequency $\Omega/2\pi$, which are modulated by fast oscillations of frequency $2\tilde f_\text{res}$, originating from the counter-rotating term.
We note that fast oscillations of $P_0$ and $P_1$ at leakage frequencies, $f=(E_L-E_0)/h, (E_L-E_1)/h$, only arise at $O[A^2]$, and are therefore much smaller.
If the leakage state becomes appreciably occupied however, such oscillations would also be observed with amplitudes of $O[A]$.
Probability oscillations calculated in this way are plotted in Fig.~\ref{Fig:HybridQubitResult}(a), keeping corrections up to higher order in $A$.
We see that the analytical results from our dressed-state theory are generally quite accurate. 
As before, we also note that although the results here were obtained for the special case of resonant driving, the general evolution operator corresponding to arbitrary frequencies can also be derived by the same method.

\begin{figure}[t]
\includegraphics[width=2.4in]{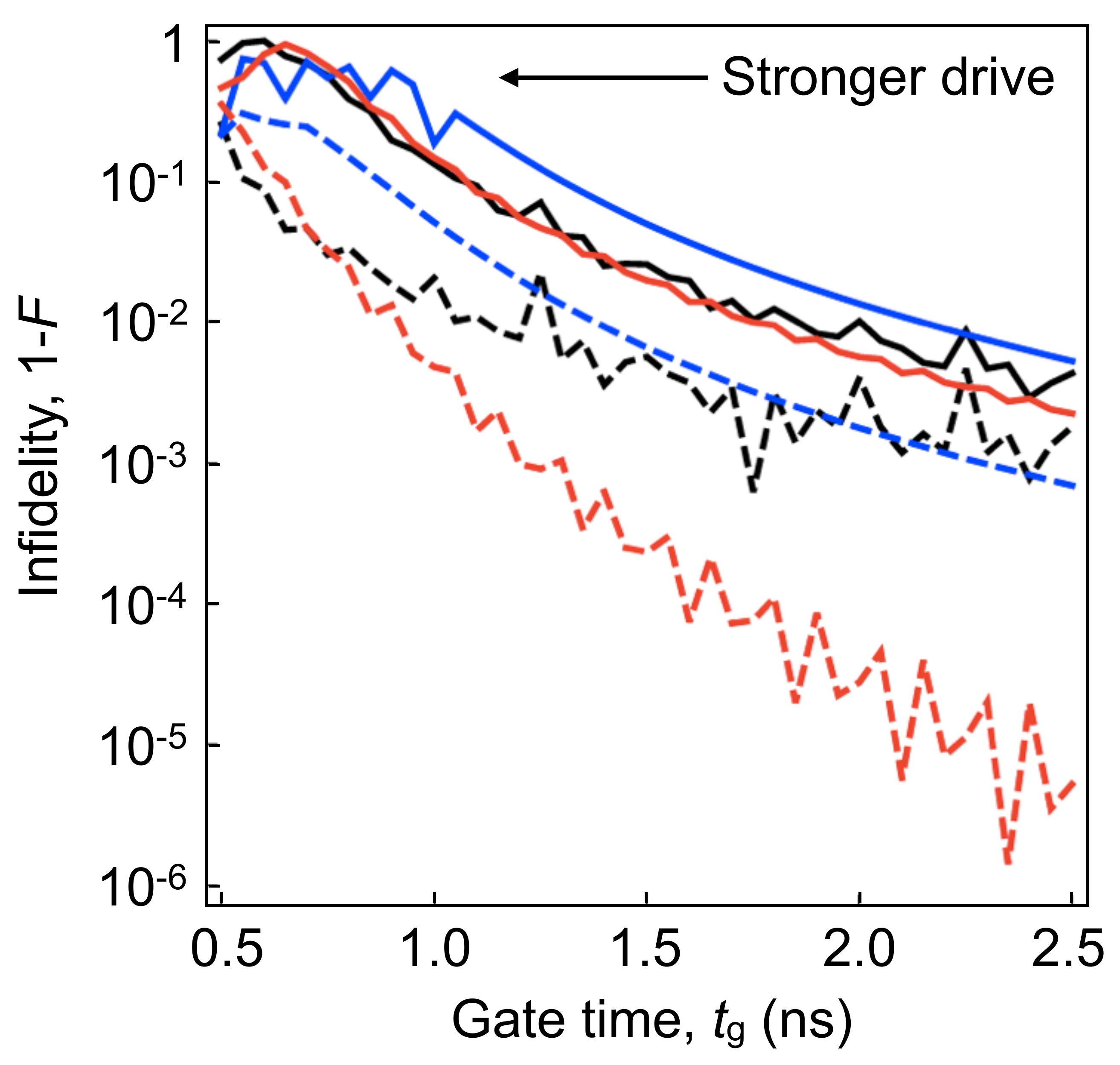}
\caption{
\label{Fig:DrivingDetuning}
Simulation results for the infidelity, $1-F$, of an $X_\pi$ rotation, obtained by driving the detuning.
The results should be compared to Fig.~\ref{Fig:Results}(c), where tunnel coupling driving was used.
In both cases, we assume the same control parameters, $E_\text{ST}/h = 12$ GHz and $\{\varepsilon,\Delta_{1},\Delta_{2}\} = \{6,0.7,0.7\} E_\text{ST}$.
The colors and line styles also have the same meaning as in Fig.~\ref{Fig:Results}(c), with one exception:  in Fig.~\ref{Fig:Results}(c), the dashed lines corresponded to including strong-driving corrections for $\tilde f_\text{res}$ and $f_\text{Rabi}$ up to $O[(A/h\! f)^2]$, while here, we include corrections up to $O[(A/h\! f)^4]$.
}
\end{figure}

From Eqs.~(\ref{eq:UL0}) and (\ref{eq:UL1}), we see that oscillations of $P_L$ occur at relatively high frequencies corresponding to leakage, as consistent with the results observed in Fig.~\ref{Fig:HybridQubitResult}(b).
This behavior can be traced back to the hybridization of leakage and logical states in Eq.~(\ref{Eq:hybridstateL}), since there are no direct couplings between leakage and logical states in Eq.~(\ref{Eq:HblockSecond}).
\red{Here, the evolution of the leakage state is fully coherent, since noise is not included in the model.}

\begin{figure}[t]
\includegraphics[width=3.2in]{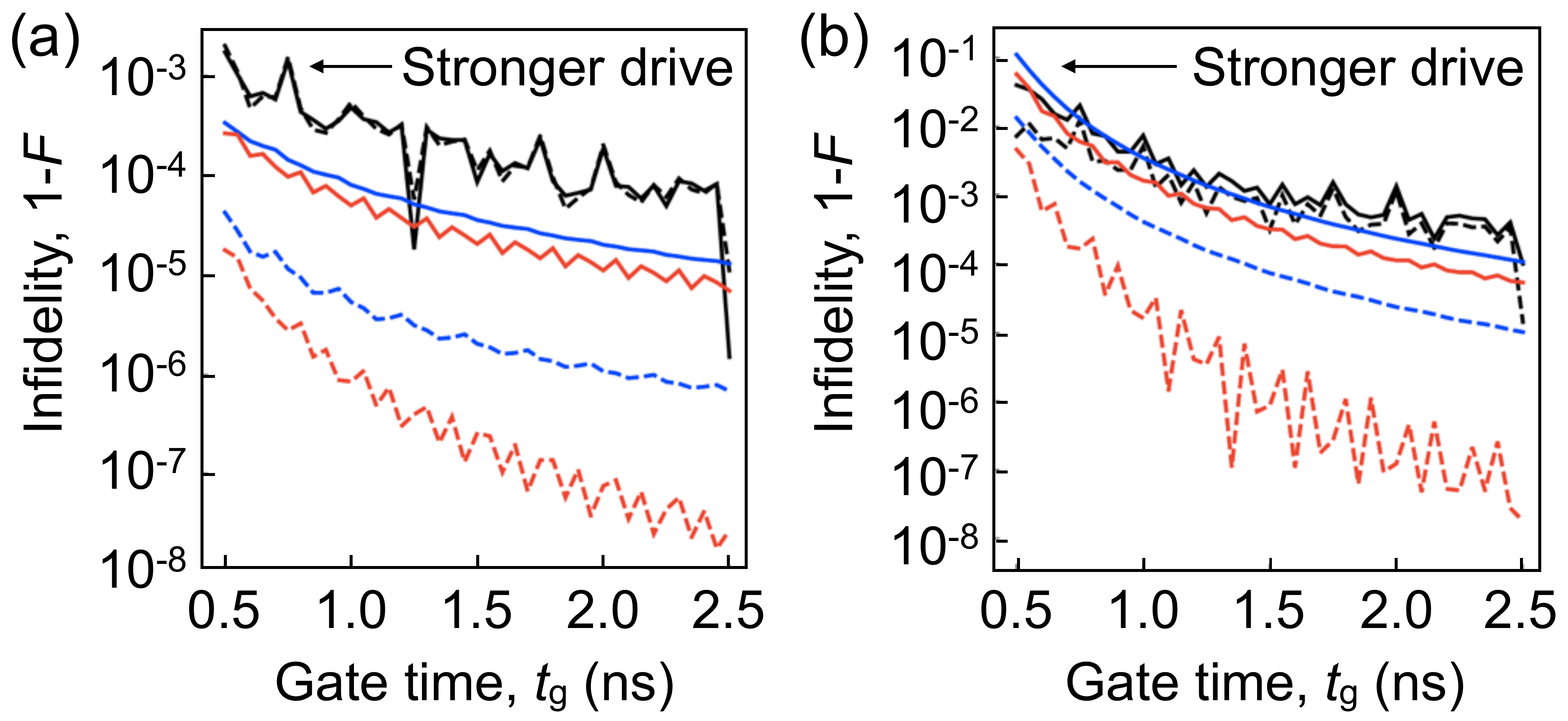}
\caption{
\label{Fig:XPI2}
Simulation results for the infidelity, $1-F$, of an $X_{\pi/2}$ rotation, using (a) tunnel coupling driving, and (b) detuning driving.
The simulation parameters used here are the same as those mentioned in Figs.~\ref{Fig:Results}(c) and \ref{Fig:DrivingDetuning}, where $X_\pi$ gates were investigated.
The colors and line styles here have the same meaning as those figures.
}
\end{figure}

\section{Detuning Driving}  \label{Sec:PulseShapeDetuning}
In Figs.~\ref{Fig:HybridQubitResult} and \ref{Fig:Results} of the main text, we plot results based on tunnel coupling driving.
In this Appendix, we perform simulations using detuning driving.
The calculations use the control parameters $E_\text{ST}/h=12$ GHz and $\{\varepsilon, \Delta_1, \Delta_2\} = \{6,0.7,0.7\} E_\text{ST}$, which are the same as those used in Fig.~\ref{Fig:Results}(c), and correspond to the large-detuning regime that is commonly used in experiments, since it affords partial protection against charge noise~\cite{Shi2012}.
Here, we use the same pulse shapes defined in Eqs.~(\ref{Eq:Rectangular})-(\ref{Eq:SmoothRectangular}). 
However, the pulse amplitudes are chosen to keep the gate times fixed for all the different simulations.
As before, we perform the simulations with or without strong-driving corrections for $f_\text{Rabi}$ and $\tilde f_\text{res}$~\cite{DetuningDrive}.
The resulting process fidelities for $X_\pi$ gates are plotted in Fig.~\ref{Fig:DrivingDetuning}.

The overall trends in Fig.~\ref{Fig:DrivingDetuning} are similar to those observed in Fig.~\ref{Fig:Results}(c).
However there are at least two important differences, which can both be attributed to the same physics:
(1) the detuning driving fidelities are typically lower (i.e., the infidelities are higher), compared to tunnel coupling driving at comparable gate times, (2) the order of the strong-driving corrections needed to achieve high-fidelity gates is higher for detuning driving ($O[(A/h\! f)^4]$), compared to tunnel coupling driving ($O[(A/h\! f)^2]$).
Both of these effects arise from the fact that the Rabi frequency scales as $(\Delta_{1,2}/\varepsilon)^2$ for detuning driving, while it scales as $(\Delta_{1,2}/\varepsilon)$ for tunnel coupling driving.  
In the large-detuning regime, detuning driving therefore causes much slower gates for the same driving amplitude.
Since gate times are held fixed in our simulations, detuning driving therefore requires us to apply larger driving amplitudes $A$ to compensate for the gate speeds.
In turn, the stronger drive induces gate errors and reduces the fidelity.
Similarly, higher-order correction terms have a stronger effect in Fig.~\ref{Fig:DrivingDetuning} due to the stronger drive.

In conclusion, for the different detuning-driving scenarios shown in Fig.~\ref{Fig:DrivingDetuning}, we find that only smoothed rectangular pulses with strong-driving corrections provide gates with relatively high fidelities at short gate times.
For a gate time of 1~ns in these simulations, the observed fidelity is better than 99\%, while for a gate time of 2~ns, the fidelity improves to 99.99\%.

\section{$X_{\pi/2}$ gates \label{Sec:XPI2}}
In this Appendix, we perform simulations to determine whether the specific choice of gate operations can affect the fidelity results.
In contrast with all other simulations up to this point, we now investigate the process fidelity of $X_{\pi/2}$ gates.
We also consider tunnel coupling driving as well as detuning driving. 

For $X_{\pi/2}$ gates, the ideal evolution operator can be expressed in the lab frame as
\begin{equation}
U_\text{ideal} = 
\begin{pmatrix}
\frac{1}{\sqrt{2}} e^{ - \frac{i}{\hbar} \tilde{E}_0 t_g} & -\frac{i}{\sqrt{2}} e^{ - \frac{i}{\hbar} \tilde{E}_0 t_g}\\
-\frac{i}{\sqrt{2}} e^{ - \frac{i}{\hbar} \tilde{E}_1 t_g} & \frac{1}{\sqrt{2}} e^{ - \frac{i}{\hbar} \tilde{E}_1 t_g}
\end{pmatrix},
\end{equation}
where $\{\tilde{E}_0,\tilde{E}_1\} = \{E_0,E_1\}$ in the RWA approximation,
and $\{\tilde{E}_0,\tilde{E}_1\} = \{E_0 + \beta_0,E_1 + \beta_1\}$ if we include strong-driving corrections.
As consistent with our previous simulations, the Bloch-Siegert shifts ($\beta_0$ and $\beta_1$) and the modified resonance and Rabi frequencies include correction terms up to $O[(A/hf)^2]$ for tunnel coupling driving, or $O[(A/hf)^4]$ for detuning driving.

The results of these $X_{\pi/2}$ simulations are shown in Fig.~\ref{Fig:XPI2}.
The general trends are similar to those observed in Figs.~\ref{Fig:Results}(c) and \ref{Fig:DrivingDetuning} for $X_\pi$ gates.
Overall, we see that fidelities are improved for $X_{\pi/2}$ gates, which can be explained by the fact that, for the same driving amplitude $A$, an $X_{\pi/2}$ gate should take about half as long as an $X_\pi$ gate.
Hence, for a fixed gate time $t_g$, the $X_{\pi/2}$ gate requires a smaller driving amplitude, which in turn improves its fidelity.
To conclude, we find that $X_{\pi/2}$ gates with $t_g=1$~ns and fidelities~$>99.99$\% can be achieved using a variety of schemes for tunnel coupling driving, but still require smoothed rectangular pulses with high-order strong-driving corrections for detuning driving.

\bibliography{DressedAtom.bib}

\begin{thebibliography}{71}
\bibitem [{Ord()}]{Orders}%
  \BibitemOpen
  \href@noop {} {}\bibinfo {note} {Since typical gate times are given by {$t_g
  \sim 1/f_\text{Rabi} \sim A^{-1}$}, if we want to include corrections in
  {$U_\text{semi}(t)$} up to {$O[(A/h\!f)^{k}]$}, then {$\tilde{
  \mathcal{H}}_\text{block}$} should include corrections up to
  {$O[(A/h\!f)^{k+1}]$} and the hybridized dressed states should include
  corrections up to {$O[(A/h\!f)^{k}]$}.}\BibitemShut {Stop}%
\bibitem [{Det()}]{DetuningDrive}%
  \BibitemOpen
  \href@noop {} {}\bibinfo {note} {When strong driving corrections are used to
  compute $f_\text{Rabi}$, the pulse amplitudes must be be rescaled to maintain
  a fixed gate time.}\BibitemShut {Stop}%
\end{thebibliography}

\end{document}